\documentclass[superscriptaddress,preprintnumbers,amsmath,amssymb,prd,nofootinbib,preprint]{revtex4-1}
\pdfoutput=1
\usepackage{graphicx}
\usepackage{epstopdf}
\usepackage{dcolumn}
\usepackage{bm}
\usepackage{hyperref}
\usepackage{color}
\usepackage{amsmath}
\usepackage{cancel}
\usepackage{xpatch}
\usepackage{mathtools}

\begin{document}


\def\a{\alpha}
\def\b{\beta}
\def\c{\varepsilon}
\def\d{\delta}
\def\e{\epsilon}
\def\f{\phi}
\def\g{\gamma}
\def\h{\theta}
\def\k{\kappa}
\def\l{\lambda}
\def\m{\mu}
\def\n{\nu}
\def\p{\psi}
\def\q{\partial}
\def\r{\rho}
\def\s{\sigma}
\def\t{\tau}
\def\u{\upsilon}
\def\v{\varphi}
\def\w{\omega}
\def\x{\xi}
\def\y{\eta}
\def\z{\zeta}
\def\D{\Delta}
\def\G{\Gamma}
\def\H{\Theta}
\def\L{\Lambda}
\def\F{\Phi}
\def\P{\Psi}
\def\S{\Sigma}

\def\o{\over}
\def\beq{\begin{align}}
\def\eeq{\end{align}}
\newcommand{\gsim}{ \mathop{}_{\textstyle \sim}^{\textstyle >} }
\newcommand{\lsim}{ \mathop{}_{\textstyle \sim}^{\textstyle <} }
\newcommand{\vev}[1]{ \left\langle {#1} \right\rangle }
\newcommand{\bra}[1]{ \langle {#1} | }
\newcommand{\ket}[1]{ | {#1} \rangle }
\newcommand{\EV}{ {\rm eV} }
\newcommand{\KEV}{ {\rm keV} }
\newcommand{\MEV}{ {\rm MeV} }
\newcommand{\GEV}{ {\rm GeV} }
\newcommand{\TEV}{ {\rm TeV} }
\newcommand{\1}{\mbox{1}\hspace{-0.25em}\mbox{l}}
\newcommand{\headline}[1]{\noindent{\bf #1}}
\def\diag{\mathop{\rm diag}\nolimits}
\def\Spin{\mathop{\rm Spin}}
\def\SO{\mathop{\rm SO}}
\def\O{\mathop{\rm O}}
\def\SU{\mathop{\rm SU}}
\def\U{\mathop{\rm U}}
\def\Sp{\mathop{\rm Sp}}
\def\SL{\mathop{\rm SL}}
\def\tr{\mathop{\rm tr}}
\def\mpl{M_{\rm Pl}}

\def\IJMP{Int.~J.~Mod.~Phys. }
\def\MPL{Mod.~Phys.~Lett. }
\def\NP{Nucl.~Phys. }
\def\PL{Phys.~Lett. }
\def\PR{Phys.~Rev. }
\def\PRL{Phys.~Rev.~Lett. }
\def\PTP{Prog.~Theor.~Phys. }
\def\ZP{Z.~Phys. }

\def\dd{\mathrm{d}}
\def\ff{\mathrm{f}}
\def\BH{{\rm BH}}
\def\inf{{\rm inf}}
\def\ev{{\rm evap}}
\def\eq{{\rm eq}}
\def\SM{{\rm sm}}
\def\Mpl{M_{\rm Pl}}
\def\GeV{{\rm GeV}}
\def\Myr{\rm Myr}
\newcommand{\Red}[1]{\textcolor{red}{#1}}
\newcommand{\TL}[1]{\textcolor{blue}{\bf TL: #1}}
\newcommand{\updm}{{\Delta'}}
\newcommand{\upex}{{h'}}

\newcommand{\DD}[1]{\textcolor{blue}{\bf DD: #1}}

\newcommand{\nocontentsline}[3]{}
\newcommand{\tocless}[2]{\bgroup\let\addcontentsline=\nocontentsline#1{#2}\egroup}
\newcommand{\toclessnonum}[2]{\bgroup\let\addcontentsline=\nocontentsline#1*{#2}\egroup}

\newcommand{\appropto}{\mathrel{\vcenter{
  \offinterlineskip\halign{\hfil$##$\cr
    \propto\cr\noalign{\kern2pt}\sim\cr\noalign{\kern-2pt}}}}}

\title{
Dark Matter, Dark Radiation and Gravitational Waves  \\
from   \\
Mirror Higgs Parity}
\author{David Dunsky}
\affiliation{Department of Physics, University of California, Berkeley, California 94720, USA}
\affiliation{Theoretical Physics Group, Lawrence Berkeley National Laboratory, Berkeley, California 94720, USA}
\author{Lawrence J. Hall}
\affiliation{Department of Physics, University of California, Berkeley, California 94720, USA}
\affiliation{Theoretical Physics Group, Lawrence Berkeley National Laboratory, Berkeley, California 94720, USA}
\author{Keisuke Harigaya}
\affiliation{School of Natural Sciences, Institute for Advanced Study, Princeton, New Jersey, 08540}

\begin{abstract}
An exact parity replicates the Standard Model giving a Mirror Standard Model, SM $\leftrightarrow$ SM$'$. This ``Higgs Parity" and the mirror electroweak symmetry are spontaneously broken by the mirror Higgs, $\vev{H'} = v' \gg \vev{H}$, yielding the Standard Model Higgs as a Pseudo-Nambu-Goldstone Boson of an approximate $SU(4)$ symmetry, with a quartic coupling $\lambda_{SM}(v') \sim 10^{-3}$. Mirror electromagnetism is unbroken and dark matter is composed of $e'$ and $\bar{e}'$. Direct detection may be possible via the kinetic mixing portal, and in unified theories this rate is correlated with the proton decay rate. With a high reheat temperature after inflation, the $e'$ dark matter abundance is determined by freeze-out followed by dilution from decays of mirror neutrinos, $\nu' \rightarrow \ell H$.  Remarkably, this requires $v' \sim (10^8 - 10^{10})$ GeV, consistent with the Higgs mass, and a Standard Model neutrino mass of $(10^{-2} - 10^{-1})$ eV, consistent with observed neutrino masses. The mirror QCD sector exhibits a first order phase transition producing gravitational waves that may be detected by future observations. Mirror glueballs decay to mirror photons giving dark radiation with $\Delta N_{\rm eff} \sim 0.03 - 0.4$. With a low reheat temperature after inflation, the $e'$ dark matter abundance is determined by freeze-in from the SM sector by either the Higgs or kinetic mixing portal.
\vspace{0.5 cm}
\begin{center}
{\it Dedicated to the memory of Ann Nelson.}
\end{center}
\end{abstract}

\date{\today}

\maketitle

\tableofcontents

\section{Introduction}
At high energy colliders, precision measurements of the electroweak symmetry breaking sector of the Standard Model (SM) have been pursued for decades, but so far there has been no discovery of any physics that would lead to a natural explanation of the weak scale.  If the SM Effective Field Theory is valid well above the weak scale, at what mass scale will it finally break down?  A possible answer has been provided by the LHC: perhaps new physics enters at the scale where the SM Higgs quartic coupling passes through zero.  For example, this new physics could be the breaking of PQ symmetry~\cite{Redi:2012ad} or of supersymmetry~\cite{Hall:2013eko,Ibe:2013rpa,Hall:2014vga,Fox:2014moa}.  

Another possibility for this new physics is the breaking of a discrete symmetry, ``Higgs Parity", that interchanges the SM Higgs, $H$, a doublet under the weak $SU(2)$, with a partner Higgs, $H'$, a doublet under some $SU(2)'$~\cite{Hall:2018let}.  There are many implementations of this idea.  One elegant possibility is that $SU(2)'$ is identified as the $SU(2)_R$ under which the right-handed quarks and leptons transform as doublets.  In this case Higgs Parity may include spacetime parity and lead to a solution of the strong CP problem~\cite{Hall:2018let}: parity forces $\theta$ to vanish and the quark Yukawa matrices to be Hermitian~\cite{Beg:1978mt,Mohapatra:1978fy,Babu:1988mw,Babu:1989rb}. Furthermore, since the breaking of $SU(3) \times SU(2)_L \times SU(2)_R \times U(1)_{B-L}$ occurs at the scale where the SM Higgs quartic vanishes, a remarkably successful unification of couplings results~\cite{Hall:2018let,Hall:2019qwx}. However, the theory needs extending to incorporate dark matter (DM).

In another class of theories, Higgs parity transforms SM quarks and leptons, $(q,u,d,l,e)$, into mirror quarks and leptons, $(q', u', d', l', e')$.  We have recently explored such a theory where the electroweak group is doubled, but QCD is not, so both ordinary and mirror quarks are colored~\cite{Dunsky:2019api}.  This theory solves the strong CP problem, with mirror quark contributions to $\bar{\theta}$ cancelling contributions from the ordinary quarks~\cite{Barr:1991qx}.  Although there is no immediate path to gauge coupling unification, the theory does have the interesting possibility of $e'$ dark matter that is within reach of direct detection.  However, hadrons containing the $u'$ quark are also stable, and since the bounds on such heavy hadron dark matter are very strong, the $e'$ production mechanism must be non-thermal rather than thermal.

In this paper we study a complete mirror sector where Higgs Parity doubles the entire Standard Model: ${\rm SM} \leftrightarrow {\rm SM}'$.   In this theory $e'$ and $u'$ are again stable and DM candidates; but since now $u'$ does not couple to QCD, it is much less constrained by direct detection, allowing successful DM production via Freeze-Out with dilution or via Freeze-In.  Long ago, a mirror copy of the SM with an unbroken parity was introduced as a way to restore space-time inversion symmetry~\cite{Lee:1956qn,Kobzarev:1966qya,Pavsic:1974rq,Foot:1991bp}.

This Mirror Higgs Parity theory is highly constrained: the parameters in the SM$'$ Lagrangian are the same as in the SM Lagrangian, so that the only new parameters are the ones describing portal interactions: one for kinetic mixing, one for the Higgs portal and several for the neutrino portal.  Although the doubling of QCD implies that Higgs Parity can no longer solve the strong CP problem, there is now a gravity wave (GW) signal from the QCD$'$ transition.  In the case of Freeze-Out DM, once the neutrino portal parameters are chosen to give the observed DM abundance, the GW signal can be computed entirely in terms of measured SM parameters.  This paper is devoted to the DM, dark radiation (DR) and GW signals and their relation.

In section~\ref{sec:Z2} we review how Higgs Parity predicts the vanishing Higgs quartic coupling at a high energy scale. Section~\ref{sec:MSM} introduces the mirror copy of the SM with Higgs Parity and the mass spectrum of the mirror sector. Direct detection of DM and, in unified theories, its relation to the proton decay rate is discussed in section~\ref{sec:DD}. The constraint from long-lived mirror glueballs is investigated in section~\ref{sec:BBNandDR}. In section~\ref{sec:cosmoAbundances}, we compute the relic abundance of $e'/u'$ dark matter and dark radiation. The spectrum of the GWs from the mirror QCD phase transition is estimated in section~\ref{sec:GW}. The final section is devoted to conclusions and discussions.

\section{Vanishing Higgs Quartic from a $Z_2$ Symmetry}
\label{sec:Z2}
In this section we review the framework of~\cite{Hall:2018let} that yields the near vanishing of the SM Higgs quartic coupling at a high energy scale. 
Consider a $Z_2$ symmetry that exchanges the $SU(2)$ weak gauge interaction with a new $SU(2)'$ gauge interaction, and the Higgs field $H(2,1)$ with its partner $H'(1,2)$, where the brackets show the $(SU(2), SU(2)')$ representation. We call the $Z_2$ symmetry as Higgs Parity. The scalar potential for $H$ and $H'$ is
\begin{align}
\label{eq:potential}
V(H,H') = - m^2 (H^\dagger H + H'^\dagger H') + \frac{\lambda}{2} (H^\dagger H + H'^\dagger H')^2 + \lambda' H^\dagger H H'^\dagger H' .
\end{align}
We assume that the mass scale $m$ is much larger than the electroweak scale. With $m^2$ positive, the Higgs parity is spontaneously broken and $H'$ acquires a large vacuum expectation value of $\vev{H'} = v'$, with $v'^2 = m^2/\lambda$. After integrating out $H'$ at tree-level, the Low Energy potential in the effective theory for $H$ is
\begin{align}
\label{eq:potentialLE}
V_{LE}(H) = \lambda' \; v'^2  \; H^\dagger H - \lambda' \left(1  + \frac{\lambda'}{2 \lambda} \right) (H^\dagger H)^2 .
\end{align}
To obtain the hierarchy $\vev{H} = v \ll v'$, it is necessary to tune $\lambda'$ to a very small value $\lambda'  \sim - v^2/v'^2$; the quartic coupling of the Higgs $H$, $\lambda_{\rm SM}$, is then extremely small.  

The vanishing quartic can be understood by an accidental $SU(4)$ symmetry under which $(H, H')$ is in a fundamental representation. For $|\lambda'| \ll 1$,  necessary for $v \ll v'$, the potential in Eq.~(\ref{eq:potential}) becomes $SU(4)$ symmetric.
After $H'$ obtains a vacuum expectation value, the SM Higgs is understood as a Nambu-Goldstone boson with a vanishing potential.
Note that in this limit of extremely small $\lambda'$, the vacuum alignment in the SU(4) space is determined by the Coleman-Weinberg potential.  The top contribution beats the gauge contribution so that the true vacuum is the asymmetric one, where the entire condensate lies in $H'$ (or in $H$, which is physically equivalent). (The $SU(4)$ symmetry implies that the Higgs boson contribution to the Coleman-Weinberg potential does not affect the vacuum orientation.)
 
Below the scale $v'$, quantum corrections from SM particles renormalize the quartic coupling, and it becomes positive.
From the perspective of running from low to high energies, the scale at which the SM Higgs quartic coupling vanishes
is identified with $v'$.
The threshold correction to $\lambda_{\rm SM}(v')$ is calculated in the next section.

Although the scale $v'$ is much smaller than the Planck scale and the typical unification scale, the theory is no more fine-tuned than the SM because of Higgs Parity.
The required fine-tuning of the theory is
\begin{align}
\frac{m^2}{\Lambda^2} \times \frac{v^2}{m^2} = \frac{v^2}{\Lambda^2},
\end{align}
where the first factor in the left hand side is the fine-tuning to obtain the scale $m$ much smaller than the cut off scale $\Lambda$, and the second one is the fine-tuning in $\lambda'$ to obtain the electroweak scale from $m$. The total tuning is the same as in the SM, $v^2 / \Lambda^2$, and may be explained by environment requirements~\cite{Agrawal:1997gf,Hall:2014dfa}.

\section{The Mirror Standard Model}
\label{sec:MSM}
The phenomenology of the theory crucially depends on the action of Higgs Parity on the SM gauge group. Refs.~\cite{Hall:2018let,Hall:2019qwx} considers the case where the $SU(3)_c \times U(1)_Y$ gauge group is not replicated. The theory solves the strong CP problem and can be embedded into $SO(10)$ unification. Ref.~\cite{Dunsky:2019api} replicates the $U(1)_Y$ gauge group. The theory solves the strong CP problem and has an interesting dark matter candidate.
In this paper we study a theory where the SM gauge group is entirely replicated by a $Z_2$ symmetry which maps
\begin{align}
\label{eq:Paction}
 SU(3) \times SU(2) \times U(1) \; \; \leftrightarrow \; \; & SU(3)' \times SU(2)' \times U(1)' \nonumber \\
 q, \bar{u}, \bar{d}, \ell, \bar{e} \; \; \leftrightarrow \; \; & q', \bar{u}', \bar{d}', \ell', \bar{e} \nonumber \\
 H \; \; \leftrightarrow \; \; & H'.
\end{align}
where matter is described by 2-component, left-handed, Weyl fields. 
\footnote{The $Z_2$ mapping described in \eqref{eq:Paction} is not unique. For example, the $Z_2$ symmetry can be extended to spacetime parity if space is inverted and SM fields are mapped to their Hermitian conjugated mirrors.}
\subsection{The Lagrangian}
The most general gauge and Higgs Parity invariant Lagrangian up to dimension $5$ is
\begin{equation}
\begin{split}
\label{eq:L}
{\cal L} \; =& \; {\cal L}_{SM} (q, \bar{u},\bar{d},l,\bar{e},H) + {\cal L}_{SM'}(q', \bar{u}',\bar{d}',l',\bar{e}',H') + \lambda'' (H^\dagger H)(H'^\dagger H') + \frac{\epsilon}{2} \; B^{\mu \nu} B'_{\mu \nu} \\
 & + \, (\ell \, \eta \, \ell) \frac{H^2}{M_M}   +  (\ell' \, \eta \, \ell') \frac{H^{'2}}{M_M}  + (\ell \, \xi \, \ell') \frac{H H'}{M_D} 
+ {\rm h.c.}
\end{split}
\end{equation}
where ${\cal L}_{SM}$ is the SM Lagrangian up to dimension $4$ and ${\cal L}_{SM'}$ its $Z_2$ mirror. The next two terms of \eqref{eq:L} link the SM and mirror sectors: $\lambda''=\lambda + \lambda'$ describes mixing between the ordinary and mirror Higgs doublets and $\epsilon$ kinetic mixing between ordinary and mirror hypercharge. The dimension 5 operators in the second line of $\eqref{eq:L}$ describe the neutrino sector. $M_{M,D}$ are large mass scales and $\eta$ and $\xi$ are $3 \times 3$ dimensionless flavor matrices. 

\subsection{The Mirror Spectrum}

\begin{figure}[tb]
\centering
\includegraphics[width=0.7\textwidth]{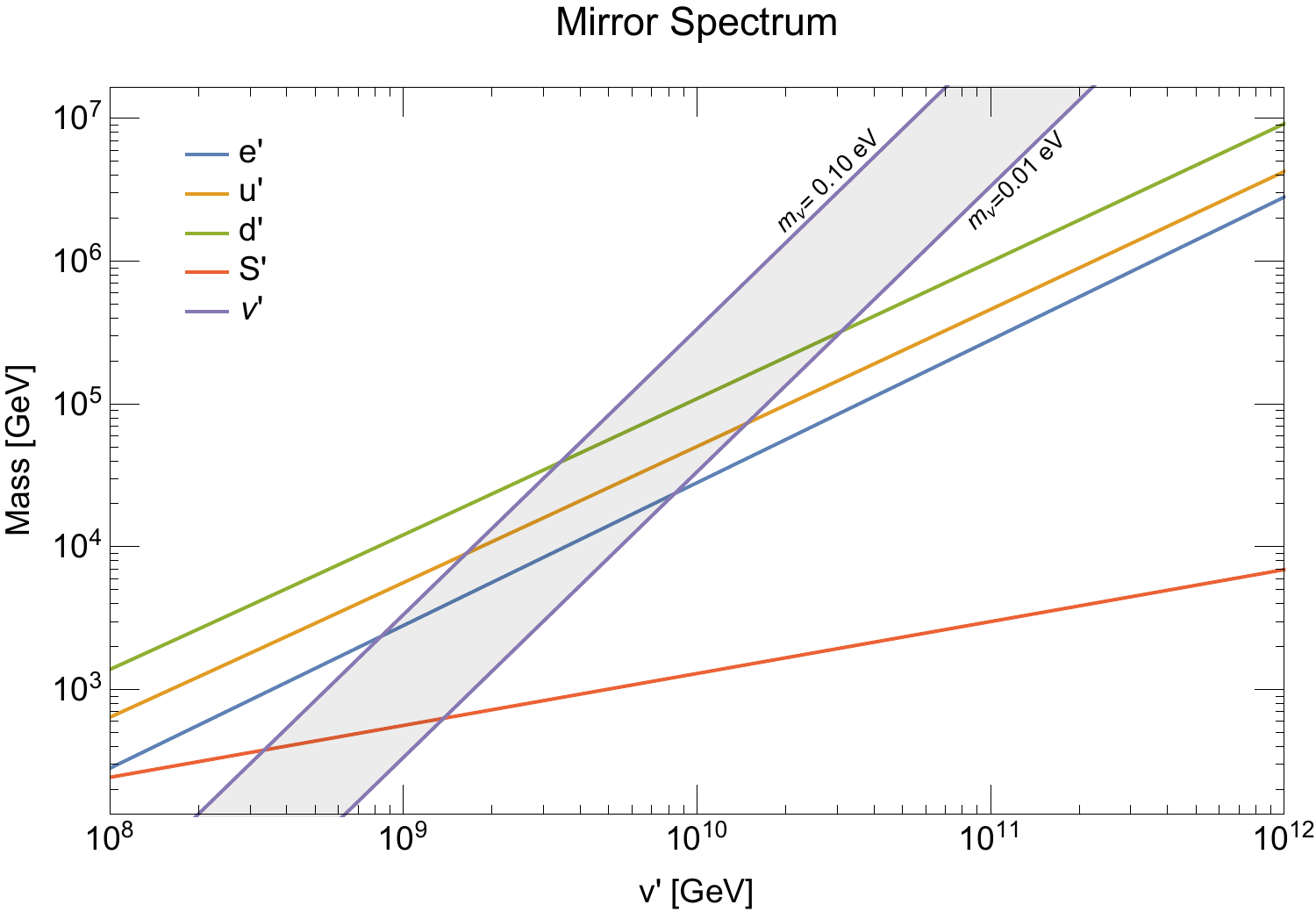} %
\caption{\small Mass spectrum of key mirror particles. The purple band shows the range of mirror neutrino masses for SM neutrino masses betwen $0.01 - 0.10 ~\rm{eV}$.}
\label{fig:mirrorSpectrum}
\end{figure}

The charged mirror fermions acquire a mass $m_{f'} = y_{f'}v'$ from the vacuum expectation value of the mirror Higgs, $v'$. The $Z_2$ symmetry sets $y_{f'} = y_f$ at the scale $\mu = v'$, so that mirror fermion masses are larger than their SM counterparts by a factor of approximately $v'/v$, as shown in Fig.~\ref{fig:mirrorSpectrum}. 

Mirror electrons and up quarks are the lightest fermions charged under $U(1)'_{EM}$ and $SU(3)'$, respectively, and thus stable and viable DM candidates. We explore $e'$ and $u'$ DM in Sec. \ref{sec:cosmoAbundances}.

Unlike mirror quarks, mirror glueballs, $S'$, acquire mass chiefly from $SU(3)'$ nonperturbative effects, with mass~\cite{Borsanyi:2012ve,Chen:2005mg}
\begin{align}
m_S' \simeq 6.8 \Lambda_{\rm{QCD}}' \gg \Lambda_{\rm{QCD}}.
\end{align}
The mirror QCD confinement scale, $\Lambda_{\rm{QCD}}'$, is not a free parameter, but is determined by running $\alpha_S({m_Z}) \simeq .1181$ up to the $Z_2$ restoration scale $v'$, equating $\alpha_S(v') = \alpha_{S'}(v')$, and then running $\alpha_{S'}$ down to lower scales until it diverges at the scale $\Lambda_{\rm{QCD}}'$.
In the $\overline{\rm MS}$ scheme the dynamical scale is given by
\begin{align}
	\Lambda_{\rm{QCD}}'  \simeq 190~\GeV \left(\frac{v'}{10^{10} ~\GeV}\right)^{4/11}.
\end{align}
Mirror glueballs are unstable and dominantly decay to $\gamma' \gamma'$, and if heavy enough, subdominantly to $H H^\dagger$. The latter are visible decays which may occur during BBN if $S'$ is long-lived. We investigate such constraints in Sec. \ref{sec:BBNandDR}. 

Standard and mirror neutrinos obtain mass from the dimension 5 operators on the second line of \eqref{eq:L}. We will be interested in small mixing between $\nu'$ and $\nu$ with $M_D \gg M_M$ so that $m_{\nu'}/m_\nu \simeq (v'/v)^2$, giving
\begin{align}
\label{eq:mnu}
m_{\nu'} \simeq 10^5 \; \GeV \left( \frac{m_\nu}{0.03 \; \mbox{eV}} \right) \left( \frac{v'}{10^{10}~{\rm GeV}} \right)^2
\end{align}
as shown in Fig.~\ref{fig:mirrorSpectrum} for two values of $m_\nu$. Mirror neutrinos are unstable and decay to $\ell H$ or if heavy enough, beta decay to $e', u', d'$. Long-lived $\nu'$ may come to dominate the energy density of the universe and release significant entropy into the SM thermal bath upon decaying. We investigate the effect of such entropy dilution on freeze-out $e'$ and $u'$ DM in Sec.~\ref{sec:fo}.

\subsection{Prediction for $v'$}

\begin{figure}[tb]
    \centering
    \begin{minipage}{0.5\textwidth}
        \centering
        \includegraphics[width=.95\textwidth]{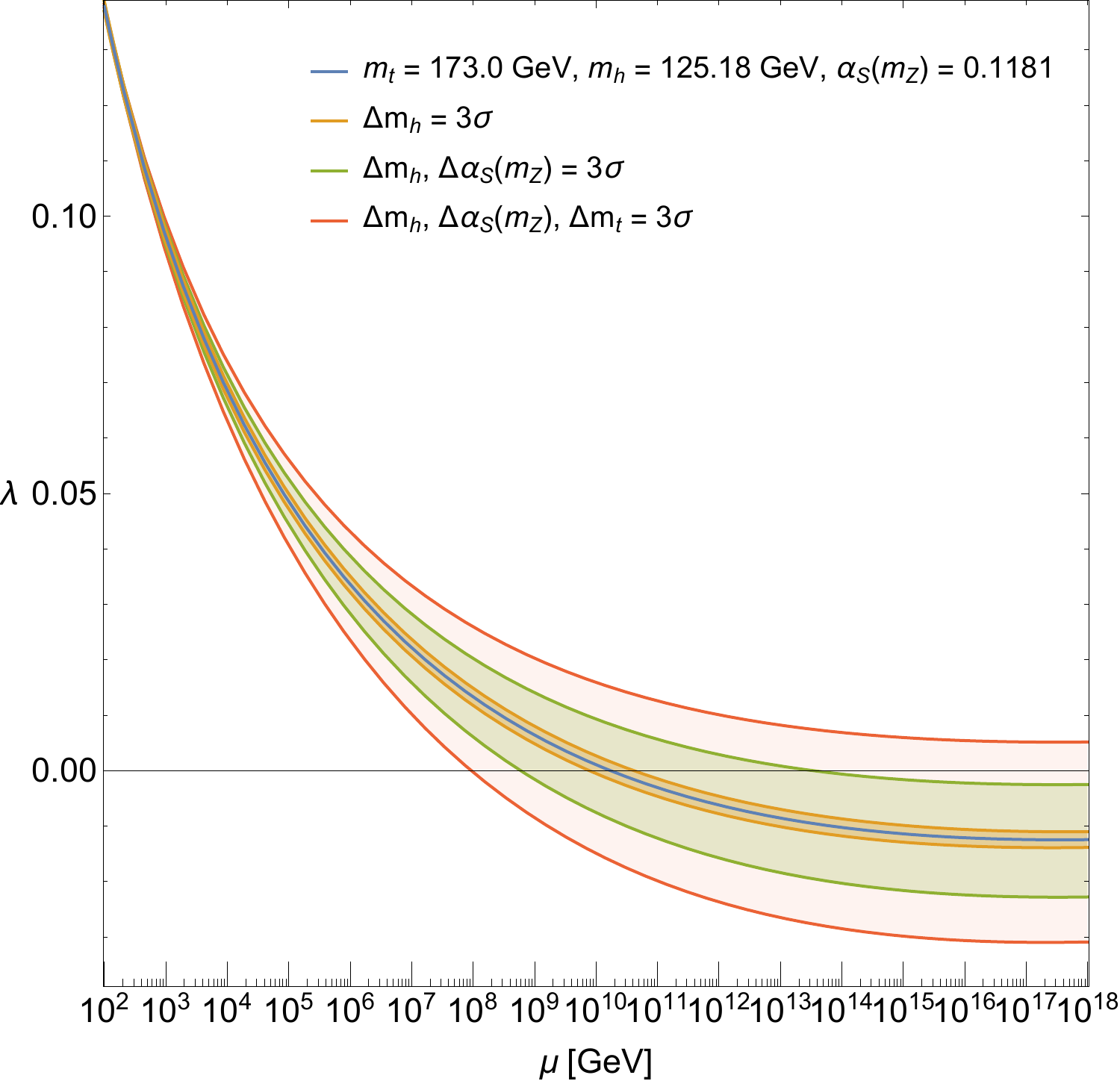} 
    \end{minipage}\hfill
    \begin{minipage}{0.5\textwidth}
        \centering
        \includegraphics[width=.95\textwidth]{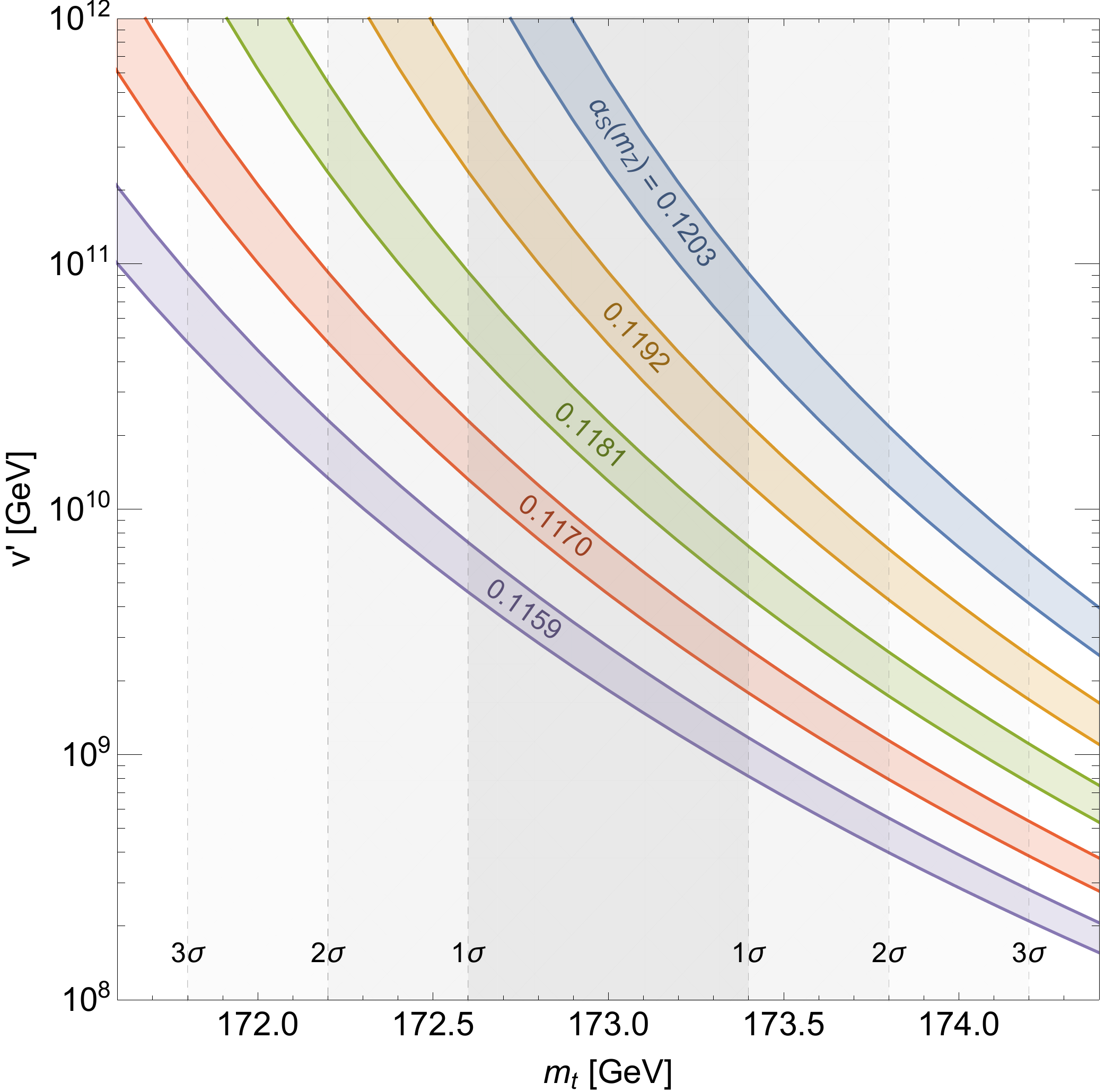} 
    \end{minipage}
    \caption{\small (Left) Running of the SM quartic coupling. (Right) Predictions for the scale $v'$ as a function of $m_t$.}
    \label{fig:vpPrediction}
\end{figure}

Between the electroweak scale and the scale $v'$, the running of the Higgs quartic coupling $\lambda_{\rm SM}$ is exactly the same as in the SM. We follow the computation in~\cite{Buttazzo:2013uya} and show the running in the left panel of Fig.~\ref{fig:vpPrediction} for a range of top quark mass $m_t = (173.0\pm 0.4)$ GeV, QCD coupling constant at the $Z$ boson mass $\alpha_S(m_Z) = (0.1181 \pm 0.0011)$, and Higgs mass $m_h = (125.18 \pm 0.16)$ GeV.

The value of the SM quartic coupling at the scale $v'$ is not exactly zero because of the threshold correction~\cite{Dunsky:2019api},
\begin{align}
\lambda_{\rm SM}(v') \simeq - \frac{3}{8\pi^2} y_t^4 \, {\rm ln} \frac{e}{y_t} + \frac{3}{128\pi^2} (g^2 + {g'}^2)^2 \, {\rm ln} \frac{e}{\sqrt{(g^2 + {g'}^2) / 2 }} + \frac{3}{64\pi^2} g^4 \, {\rm ln} \frac{e}{ g/ \sqrt{2}},
\end{align}
where the $\overline{\rm MS}$ scheme is assumed.
The prediction for the scale $v'$ is shown in the right panel of Fig.~\ref{fig:vpPrediction}. For each top quark mass and QCD coupling constant, the range of the prediction corresponds to the 1-sigma uncertainty in the measured Higgs mass, $m_h = (125.18 \pm 0.16)$ GeV. Within the uncertainties, $v'$ as small as few $10^8$ GeV is possible. Future measurements can pin down the scale $v'$ with an accuracy of few tens percent~\cite{Dunsky:2019api}.

\subsection{Kinetic Mixing}
\label{sec:kineticMixing}
\begin{figure}[tb]
\centering
\includegraphics[width=0.9\textwidth]{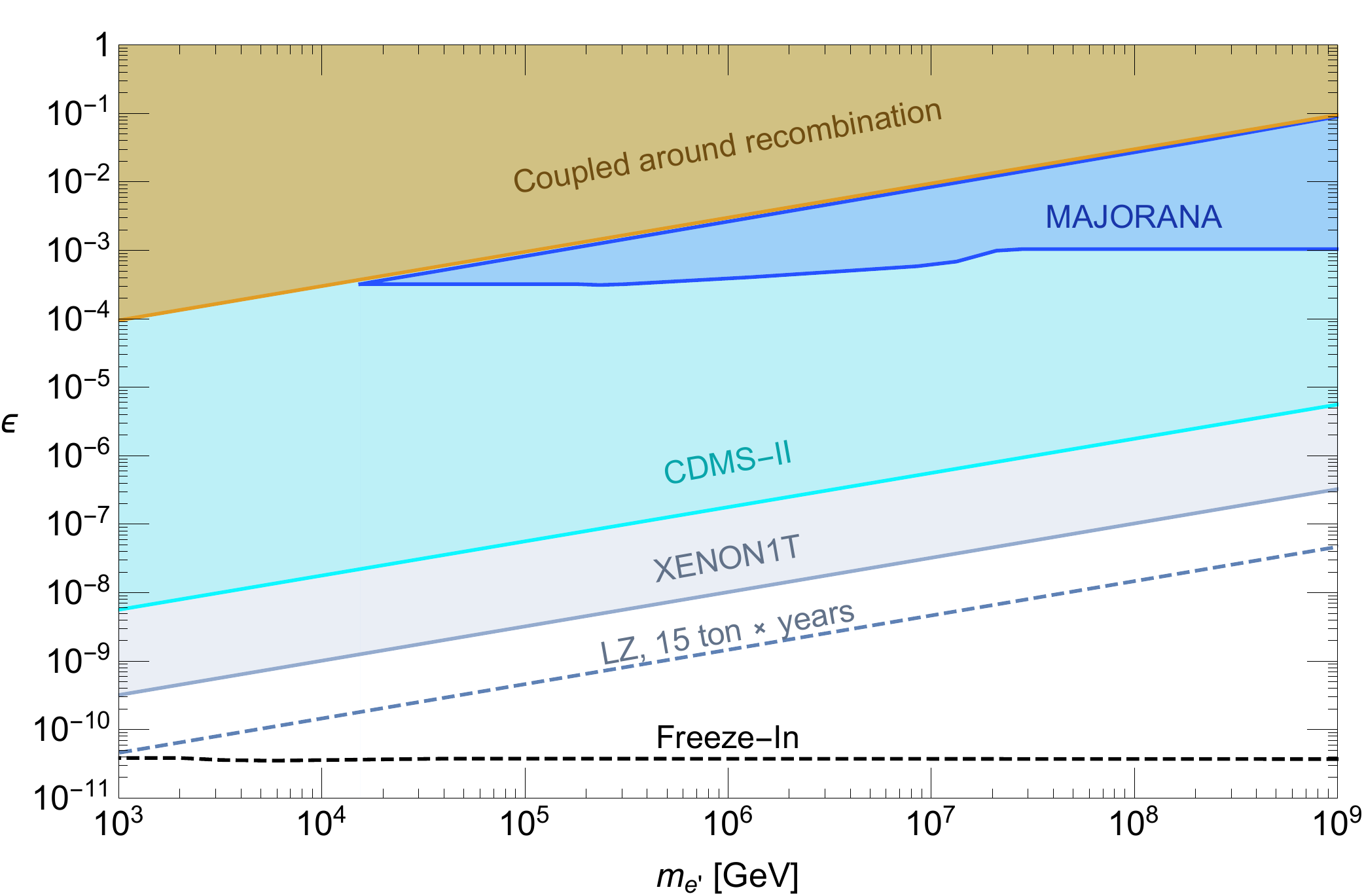} %
\caption{\small Constraints on kinetic mixing if DM is composed of mirror electrons.}
\label{fig:epsilonConstraintPlot}
\end{figure}

Even though quantum corrections to the kinetic mixing are small, 
\footnote{Diagrams contributing to kinetic mixing via the Higgs portal only occur beyond four loops, likely inducing an $\epsilon \ll 10^{-12}$.}
no symmetry forbids a tree-level $\epsilon$ from being order unity in the effective Lagrangian \eqref{eq:L}. However, as shown in Fig.~\ref{fig:epsilonConstraintPlot}, mirror electron DM with $\epsilon \gtrsim 10^{-8}$ is strongly constrained by nuclear and electron recoil experiments, ionization signals, and cosmology (\cite{Dunsky:2018mqs} and references therein.) 
A natural explanation for such a small $\epsilon$ is that $SU(3) \times SU(2) \times U(1) \times SU(3)' \times SU(2)' \times U(1)'$ unifies into a larger gauge group with no abelian factors. Consequently, $\epsilon$ must vanish above the unification scale $v_{G}$ by gauge invariance.

\begin{figure}[tb]
\centering
\includegraphics[width=0.8\textwidth]{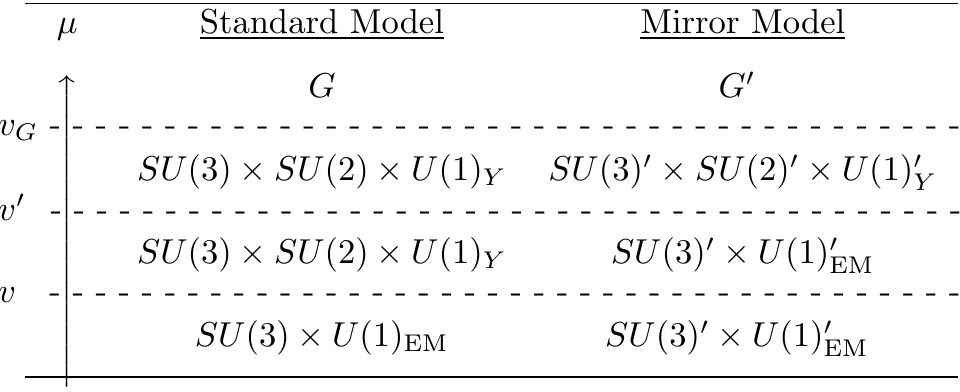} %
\caption{\small Qualitative picture of the effective field theory at scales $v, v'$, and $v_{G}$. The gauge groups $G$ and $G'$ do not contain any abelian factors so that kinetic mixing can only be radiatively generated at the scale $v_G$ and below, or be induced by higher dimensional operators at $v_G$.}
\label{fig:EFT}
\end{figure}

For example, consider a theory where the SM gauge group and the mirror gauge group separately unify to $G \times G'$ at scale $v_{G}$, shown qualitatively in Fig.~\ref{fig:EFT}. Above $v_{G}$ the operators that induce kinetic mixing between the standard and mirror sectors are:
\begin{align}
	\frac{1}{2}\frac{c_6}{M_{\rm{Pl}}^2}(\Sigma F)(\Sigma' F') + \frac{1}{2}\frac{c_8}{M_{\rm{Pl}}^4} (\Sigma^2 F)(\Sigma'^2 F') + \mathcal{O}(1/M_{\rm{Pl}}^6)
	\label{eq:GUTkineticmixing}
\end{align}
where $F,F'$ are the gauge field strengths and $\Sigma,\Sigma'$ the Higgs fields. 
The first term is absent if $\Sigma$ is not an adjoint representation of $G$ or charged under some symmetry.
When $\Sigma$ and $\Sigma'$ acquire a vacuum expectation value $v_{G}$,
\footnote{Since the $Z_2$ symmetry is unbroken above $v'$, $\langle \Sigma \rangle = \langle \Sigma' \rangle = v_{G}$.}
the higher dimensional operators in \eqref{eq:GUTkineticmixing} induce a kinetic mixing $\epsilon$
\begin{align}
	\epsilon \simeq 3.5 \times 10^{-5} \, c_6 \left(\frac{v_{G}}{10^{16} ~\GeV}\right)^2 + 6.0 \times 10^{-10} \, c_8 \left(\frac{v_{G}}{10^{16} ~\GeV}\right)^4 + \mathcal{O}{(v_G^6/M_{\rm{Pl}}^6)}. 
	\label{eq:eps1}
\end{align}
It is possible to freeze-in $e'$ as DM via the induced kinetic mixing of \eqref{eq:eps1}. As shown in Fig.~\ref{fig:EFT}, the correct DM abundance can be produced for a kinetic mixing parameter $\epsilon \simeq 4 \times 10^{-11}$, essentially independent of DM mass. If the dim-6 coefficient $c_6$ is non-zero, the correct $e'$ DM abundance can be produced for the unification scale $v_{G} \simeq  1 \times 10^{13} \,{c_6}^{-1/2} ~ \GeV$. If $c_6$ vanishes, and the dim-8 coefficient $c_8$ is non-zero, the correct $e'$ DM abundance can be produced for $v_G \simeq 5 \times 10^{15} \,{c_8}^{-1/4} ~ \GeV$.

\section{Direct detection and the Correlation with Proton Decay}
\label{sec:DD}

\subsection{Direct Detection by Nuclear Recoils}
Kinetic mixing induced from higher dimensional operators allows $e'$ dark matter to scatter electromagnetically with a nucleus. The Rutherford cross section for scattering between $e'$ and a nucleus of mass $m_N$ and atomic number $Z$, with relative velocity $v_{\rm rel}$ is given by
\begin{align}
\frac{d \sigma} {dq} = \frac{8\pi \alpha^2 Z^2 \epsilon^2}{v_{\rm rel}^2 q^3} |F(q)|^2,
\end{align}
where $q$ is the momentum transfer and $F(q)$ is the nuclear form factor. 
The number of expected events in a direct detection experiment with an energy threshold $E_{\rm th}$, a total target mass $M_{\rm tar}$, an exposure time $T$, and atomic weight $A$ is 
\begin{align}
N_{\rm event} = 1.6 \times  \left( \frac{\epsilon}{ 10^{-8}} \right)^2 \frac{10^7 \, {\rm GeV}}{m_{e'}}  \left( \frac{Z}{54} \right)^2 \left( \frac{131}{A} \right)^2 \frac{10{\rm \, keV}}{E_{\rm th}} \frac{f(E_{\rm th})}{0.3} \frac{M_{\rm tar} t}{{\rm ton}\times {\rm year}},
\label{eq:Nevent}
\end{align}
where we assume a local DM density of $0.3$ GeV/cm$^3$, as well as a velocity distribution of
\begin{align}
d v f(v) = dv  \frac{4}{\sqrt{\pi}} \frac{v^2}{v_0^3} \, {\rm exp}(- v^2/v_0^2),~~v_0= 220 \, {\rm km}/{\rm s}.
\end{align}
Here $f(E_{\rm th})$ takes into account the suppression of the scattering by the form factor,
\begin{align}
f(E_{\rm th}) = \left[ \int_{q_{\rm th}}^{q_{\rm max}} d q |F(q)|^2 q^{-3} \right] / \left[  \int_{q_{\rm th}}^{q_{\rm max}} d q q^{-3} \right], \nonumber \\
q_{\rm th} = \sqrt{2 m_N E_{\rm th}},~ q_{\rm max}= 2 m_N v_{\rm rel}.
\end{align}
Assuming the Helm form factor~\cite{Helm:1956zz,Lewin:1995rx}, we find $f(E_{\rm th})\simeq 0.3$.

XENON1T searches for a recoil between DM and Xenon with a threshold energy around 10 keV~\cite{Aprile:2018dbl}.
The bound obtained there can be interpreted  as an upper bound of 16 on the expected number of the events. Currently, the strongest bound on $\epsilon$ for $m_{e'} > 10^2 ~\GeV$ comes from XENON1T~\cite{Dunsky:2018mqs}, requiring
\begin{align}
	\epsilon < 1 \times 10^{-10} \, \left( \frac{m_{e'}}{10^2 ~\GeV} \right)^{1/2}
\end{align}
as shown in Fig.~\ref{fig:epsilonConstraintPlot}. If $\epsilon$ is close to this bound, future experiments may detect $e'$ dark matter.
\subsection{Correlation Between Proton Decay and Direct Detection}
\begin{figure}[tb]
\centering
\includegraphics[width=0.7\textwidth]{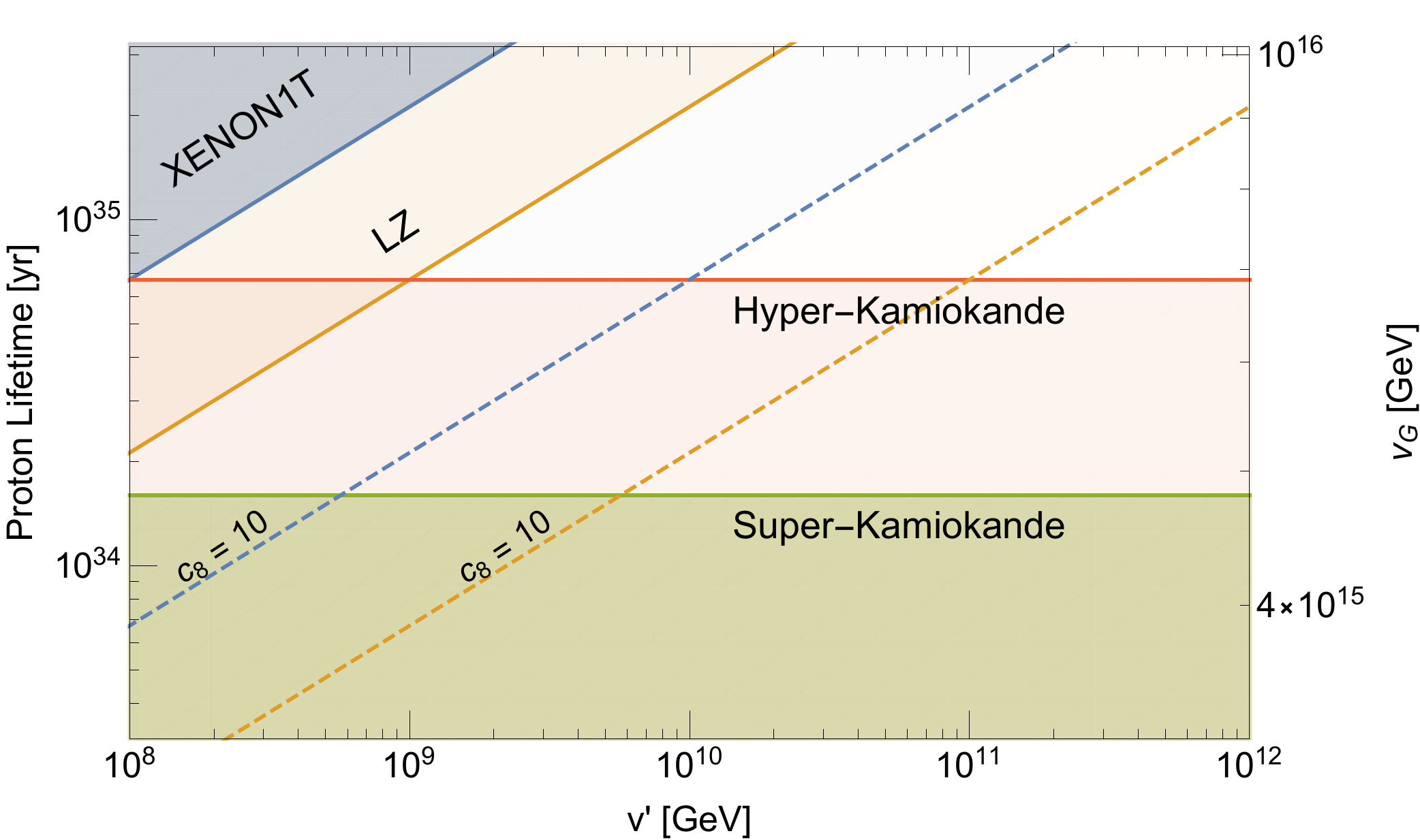} %
\caption{\small Correlation between the proton decay rate and the DM-nuclear scattering rate as a function of $v'$. The rates are related as they both depend on the unification scale $v_G$ via higher-dimensional operators.}
\label{fig:ddProtonDecay}
\end{figure}

Let us consider a case where the SM gauge group is embedded into a unified gauge group with heavy gauge bosons mediating proton decay. The proton decay rate is
\begin{align}
\Gamma^{-1}(p \rightarrow \pi^0 e^+) \simeq 3 \times 10^{35}{\rm years} \; \left( \frac{v_G}{10^{16}~{\rm GeV}} \right)^4 \left(\frac{0.103 ~\GeV^2}{W_0}\right)^2,
\end{align}
where $|W_0| = 0.103 \pm  0.041~\GeV^2$ encodes the relevant hadronic matrix element extracted from a lattice computation~\cite{Aoki:2013yxa}.
We also assume that below the heavy gauge boson mass scale the gauge group contains a $U(1)$ factor which eventually joins the $U(1)_Y$ gauge group. (This case excludes, for example, the Pati-Salam gauge group breaking at an intermediate scale.) The kinetic mixing is given by Eq.~\eqref{eq:eps1} and we assume $c_6=0$. The direct detection rate $N_{\rm event}/M_{\rm tar}t$ of (\ref{eq:Nevent}) and the proton decay rate are correlated with each other,
\begin{align}
\Gamma^{-1}(p \rightarrow \pi^0 e^+) \simeq 3 \times 10^{35} {\rm years} \; \left( \frac{N_{\rm event}}{M_{\rm tar} t} \frac{{\rm ton} \times {\rm year}}{10} \right)^{1/2} \frac{1}{c_8} \left( \frac{v'}{2 \times 10^{9}~{\rm GeV}} \right)^{1/2},
\end{align}
as shown in Fig.~\ref{fig:ddProtonDecay}. The blue region shows that if XENON1T were to detect a nuclear recoil signal, the proton lifetime would generally be longer than Hyper-Kamiokande could detect, for $c_8 = 1$. The orange region shows the analgous signal region for LZ. For $v' \leq  10^9 ~\GeV$, Hyper-Kamiokande and LZ both can detect correlating proton decay and nuclear recoil signals, respectively. If $c_8 > 1$, the kinetic mixing parameter is stronger for fixed $v_G$ so that nuclear recoil experiments and proton detect experiments may find correlating signals for $v' \gtrsim 10^9 ~\GeV$. For example, the dashed blue and orange contours of Fig.~\ref{fig:ddProtonDecay} show the reach of XENON1T and LZ, respectively, for $c_8 = 10$.

\section{High and Low Reheat Scenarios;  BBN and Dark Radiation}
\label{sec:BBNandDR}

Since all the parameters of the SM have been determined, the only free parameters that affect the cosmology of the Mirror Higgs Parity theory are the reheat temperature after inflation and the portal parameters that connect the SM and mirror sectors.  A key question is whether the two sectors were brought into thermal equilibrium after inflation.  

 At sufficiently high temperatures, the SM and mirror sectors are kept in thermal equilibrium by the Higgs portal; the sectors then decouple at a temperature 
\begin{align}
\label{eq:Tdec}
    \frac{ T_{\rm dec} }{ v' } \simeq 10^{-3} \left(\frac{v'}{10^{9} ~\GeV}\right)^{1/3}.
\end{align}
Our two cosmological scenarios correspond to whether the reheat temperature after inflation, $T_{\rm{RH}}$  is above or below $T_{\rm dec}$, and lead to very different mechanisms for the abundance of $e'$ and $u'$ dark matter.  For $T_{\rm{RH}} > T_{\rm dec}$, the $u'$ and $e'$ abundances are given by freeze-out as the temperature drops below their masses, followed by dilution from $\nu'$ decay; for $T_{\rm{RH}} < T_{\rm dec}$ we assume that only the SM sector is reheated, so that DM arises from freeze-in.  
These two schemes for DM production are discussed in the next section.

In both high and low reheating cosmologies, long-lived mirror glueballs are produced whose decay products may yield substantial dark radiation or alter the relic abundances of light elements. In this section we study the general constraints on the maximum production of mirror glueballs.  These results will be used in the next section to place limits on the high $T_{\rm{RH}}$ scheme and identify regions of parameter space that give signals of dark radiation and perturbed light element abundances.

The mirror QCD confinement transition occurs when the mirror thermal bath cools to a temperature $T'_c = 1.26 \, \Lambda'_{\rm{QCD}}$~\cite{Borsanyi:2012ve}. At this point, the mirror bath contains only $\gamma'$ and $g'$ so that the ratio of entropies of the two sectors at $T'_c$ is about $r = (16/106.75)(T'_c/T_c)^3$. If the reheat temperature after inflation is greater than $T_{\rm{dec}}$, the two sectors were initially in thermal equilibrium and $r = (8/9)(g'_*(T_{\rm{dec}})/106.75)$. On the other hand, if the reheat temperature after inflation is below $T_{\rm{dec}}$, the two sectors were never in thermal equilibrium and ratio of temperatures $T'/T$ is generally much less than one. 
\begin{figure}[tb]
    \centering
    \begin{minipage}{0.445\textwidth}
        \centering
        \includegraphics[width=1\textwidth]{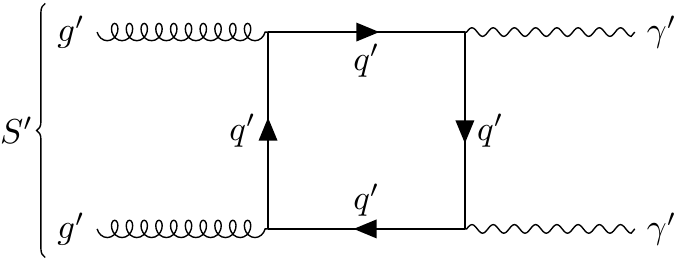} 
    \end{minipage}\hfill
    \begin{minipage}{0.455\textwidth}
        \centering
        \includegraphics[width=1\textwidth]{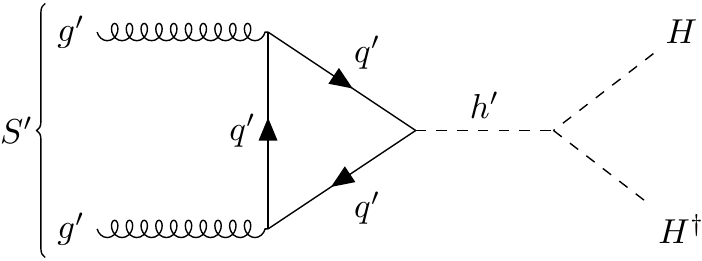} 
    \end{minipage}
    \caption{\small Mirror glueball decay to $\gamma' \gamma'$ (left) and $H, H^\dagger$ (right).}
        \label{fig:glueballDecay}
\end{figure}

After the mirror QCD transition, $g'$ confine to form mirror glueballs $S'$, whose energy density normalized by the entropy is given by
\begin{align}
    \frac{\rho_{S'}}{s} =\frac{3}{4} \, A \, r \, T'_c.
\end{align}
The factor $A$ takes into account the non-trivial dynamics before and after the phase transition and is estimated in Appendix~\ref{sec:rhos}. $A=1$ corresponds to the limit where massless ideal gas of mirror gluons suddenly becomes pressure-less mirror glueballs at $T_c'$ and the mirror glueball number density is conserved afterward. 

Mirror glueballs are typically long-lived. The lifetime of the mirror glueball is dominantly set by its decay rate to mirror photons, described by the dimension-8 operator $F'F' G'G'$,  generated by a loop of mirror quarks of mass $m_{q'}$ and charge $Q'$ as shown in the left panel of Fig.~\ref{fig:glueballDecay}.  After confinement this becomes a dimension-5 operator connecting $S'$ to $\gamma' \gamma'$~\cite{Juknevich:2009ji}
\begin{align}
    \Delta \mathcal{L}_{S' \to \gamma'\gamma'} \, &=  \, \frac{Q'^2}{240\pi} \,  \frac{ \alpha}{m_{q'}^4} \; \mathbf{F_{0^{++}}^{S'}} \; \; F'_{\mu \nu}F'^{\mu \nu} S'
\end{align}
with matrix element $\mathbf{F_{0^{++}}^{S'}} = \langle 0 | 1/2 g_s^2 G^a_{\mu \nu} G_a^{\mu\nu} | 0^{++} \rangle \simeq 2.7 m_{S'}^3$~\cite{Meyer:2008tr}.  Since the amplitude is dominated by the smallest $m_{q'}$, we take $q' = u'$ giving $Q' = 2/3$, so that the  mirror glueball decay rate to mirror photons is
\begin{align}
	\Gamma_{S' \to \gamma' \gamma'} \, \simeq \, \frac{1}{16\pi}\left(\frac{2.7 \alpha}{270 \pi } \right)^2 \frac{m_{S'}^9}{m_{u'}^8}.
\end{align}
The mirror glueball can also decay to the SM sector via the Higgs portal as shown by the right panel of Fig.~\ref{fig:glueballDecay}.  The decay rate to $H H^\dagger$ is given by
\begin{align}
	\Gamma_{S' \to H H^\dagger} \, \simeq \, \frac{1}{8\pi}\left(\frac{2.7}{16 \pi^2}\right)^2 \frac{m_{S'}^5}{v'^4}.
\end{align}
If its lifetime, $\Gamma_{S'}^{-1} \simeq (\Gamma_{S' \to \gamma' \gamma'} + \Gamma_{S' \to H H^\dagger})^{-1}$, exceeds about $1~\rm{s}$, $S'$ decays during BBN. If this occurs, $S'$ may inject substantial energy density, $\rho_{vis}$, into the SM hadronic sector altering the neutron to proton ratio before nucleosynthesis or disassociating light elements immediately after, leading to the constraint~\cite{Kawasaki:2004qu}
\begin{align}
    \frac{\rho_{vis}}{s} = \frac{\Gamma_{S' \to H H^\dagger}}{\Gamma_{S'}} \, \frac{3}{4} \, A  \,\frac{r}{D} \, T'_c \; \lesssim \; 10^{-14} ~\GeV.
\end{align}
Here, $D$ is a generic dilution factor which may arise if there exists a particle which comes to dominate the energy density of the universe and decays before BBN, thereby injecting entropy into the SM thermal bath. 

\begin{figure}[tb]
    \centering
    \begin{minipage}{0.5\textwidth}
        \centering
        \includegraphics[width=.95\textwidth]{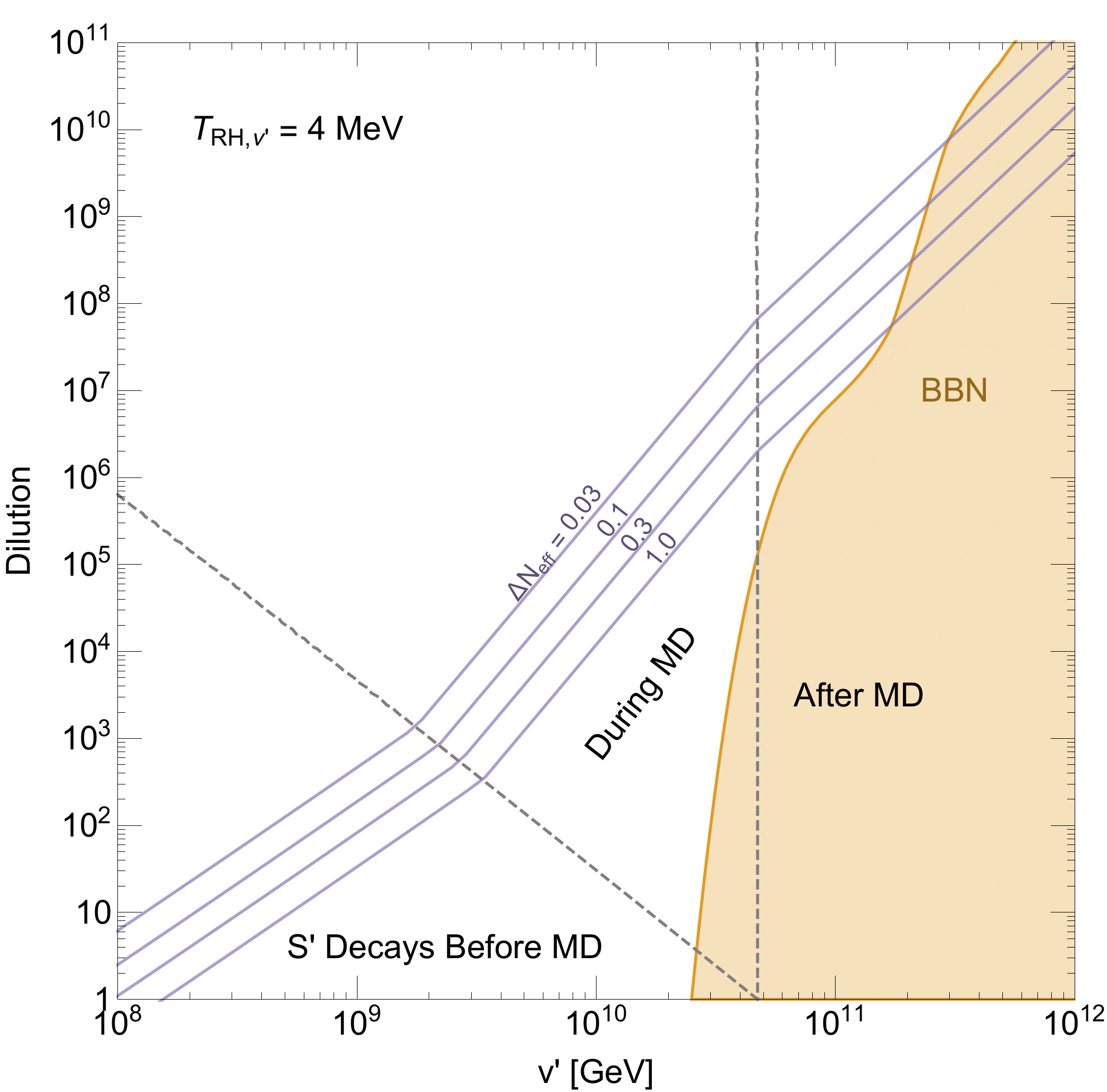} 
    \end{minipage}\hfill
    \begin{minipage}{0.5\textwidth}
        \centering
        \includegraphics[width=.95\textwidth]{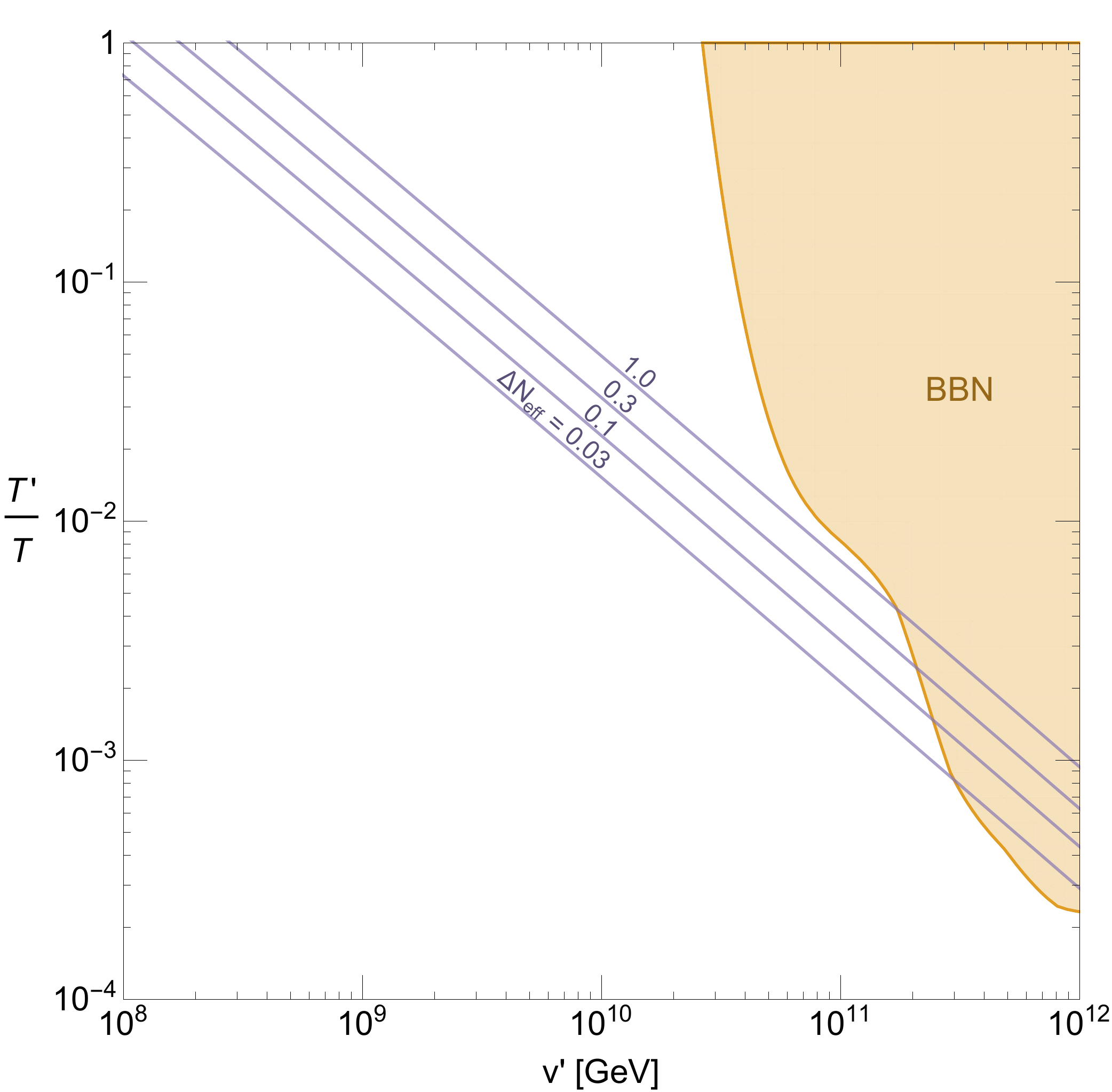} 
    \end{minipage}
    \caption{\small $\Delta N_{\rm{eff}}$ contours (purple) and BBN constraints (orange) from $S' \to \gamma' \gamma', HH^\dag$.  In the left (right) panel the two sectors were (were not) initially thermally coupled so that DM is thermally produced via freeze-out and dilution (freeze-in).  The temperature ratio of the two sectors, $T'/T$, is evaluated at the mirror confinement temperature. For clarity, we take $A = 1$.}
    \label{fig:glueballConstraints}
\end{figure}

In the cosmology with $T_{\rm{RH}} > T_{\rm{dec}}$, mirror neutrinos are a natural candidate to provide such dilution since they are abundantly produced, decouple from the mirror bath while relativistic, and are long-lived. In this scenario, $D = T_{\rm{MD}, \nu'}/T_{\rm{RH},\nu'}$, where $T_{\rm{MD}, \nu'}$ is the temperature of the SM bath when $\nu'$ induced matter-domination begins and $T_{\rm{RH},\nu'}$ when it ends. If $T_{\rm{RH}} < T_{\rm{dec}}$, there is no particle in the mirror standard model to provide such dilution and $D = 1$.
We show the BBN constraints as a function of $v'$ in Fig.~\ref{fig:glueballConstraints} in orange using the precise energy yield constraints calculated in~\cite{Kawasaki:2004qu}. When $T_{\rm{RH}} > T_{\rm{dec}}$, $r$ is known so $D$ is constrained, as shown in the left panel of Fig.~\ref{fig:glueballConstraints}. When $T_{\rm{RH}} < T_{\rm{dec}}$, $D$ is known so $r$ is constrained, as shown in the right panel of Fig.~\ref{fig:glueballConstraints}.

In addition, the energy deposited by $S'$ into mirror photons is constrained, even if $S'$ does not decay during BBN.
The mirror photons behave as dark radiation, whose energy density is conventionally expressed as an excess in the effective number of neutrino $\Delta N_{\rm{eff}}$.
For the high $T_{\rm{RH}}$ cosmology, with $\nu'$ decay leading to a dilution factor $D$, $\Delta N_{\rm{eff}}$ depends on whether $S'$ decays before, during, or after the $\nu'$ matter-dominated era 
\begin{equation}
\begin{split}
\Delta N_{\rm{eff}} &\simeq  \frac{\Gamma_{S' \to \gamma' \gamma'}}{\Gamma_{S'}} \frac{4}{7}\left(\frac{43}{4}\right)^{4/3}\frac{r}{D} \frac{T'_c}{\sqrt{\Gamma_{S'}M_{\rm{Pl}}}} A
\\& \times
\begin{dcases}
\left(\frac{\pi^2}{10}\right)^{1/4}\frac{g_*(T_{\Gamma_{S'}})^{1/4}}{g_{*S}(T_{\Gamma_{S'}})^{1/3}} \frac{1}{D^{1/3}} & S' \text{ decays before MD}\\
\left(\frac{\pi^2}{10}\right)^{1/3} \left(\frac{T_{\rm{RH},\nu'}}{\sqrt{\Gamma_{S'}M_{\rm{Pl}}}}\right)^{1/3} & S' \text{ decays during MD} \\
\left(\frac{\pi^2}{10}\right)^{1/4}\frac{g_*(T_{\Gamma_{S'}})^{1/4}}{g_{*S}(T_{\Gamma_{S'}})^{1/3}} & S' \text{ decays after MD.}
\end{dcases}
\label{eq:Ycrit}
\end{split}
\end{equation}
For the low $T_{\rm{RH}}$ cosmology few $\nu'$ are produced, so they do not give a matter dominated era and $D=1$.  Contours of the dark radiation abundance produced from $S' \to \gamma' \gamma'$ are shown in Fig.~\ref{fig:glueballConstraints}. 
\section{Cosmological Abundance of Mirror Dark Matter}
\label{sec:cosmoAbundances}

\subsection{Freeze-Out and Dilution from $\nu'$ Decay}
\label{sec:fo}
\begin{figure}[tb]
\centering
\includegraphics[width=0.8\textwidth]{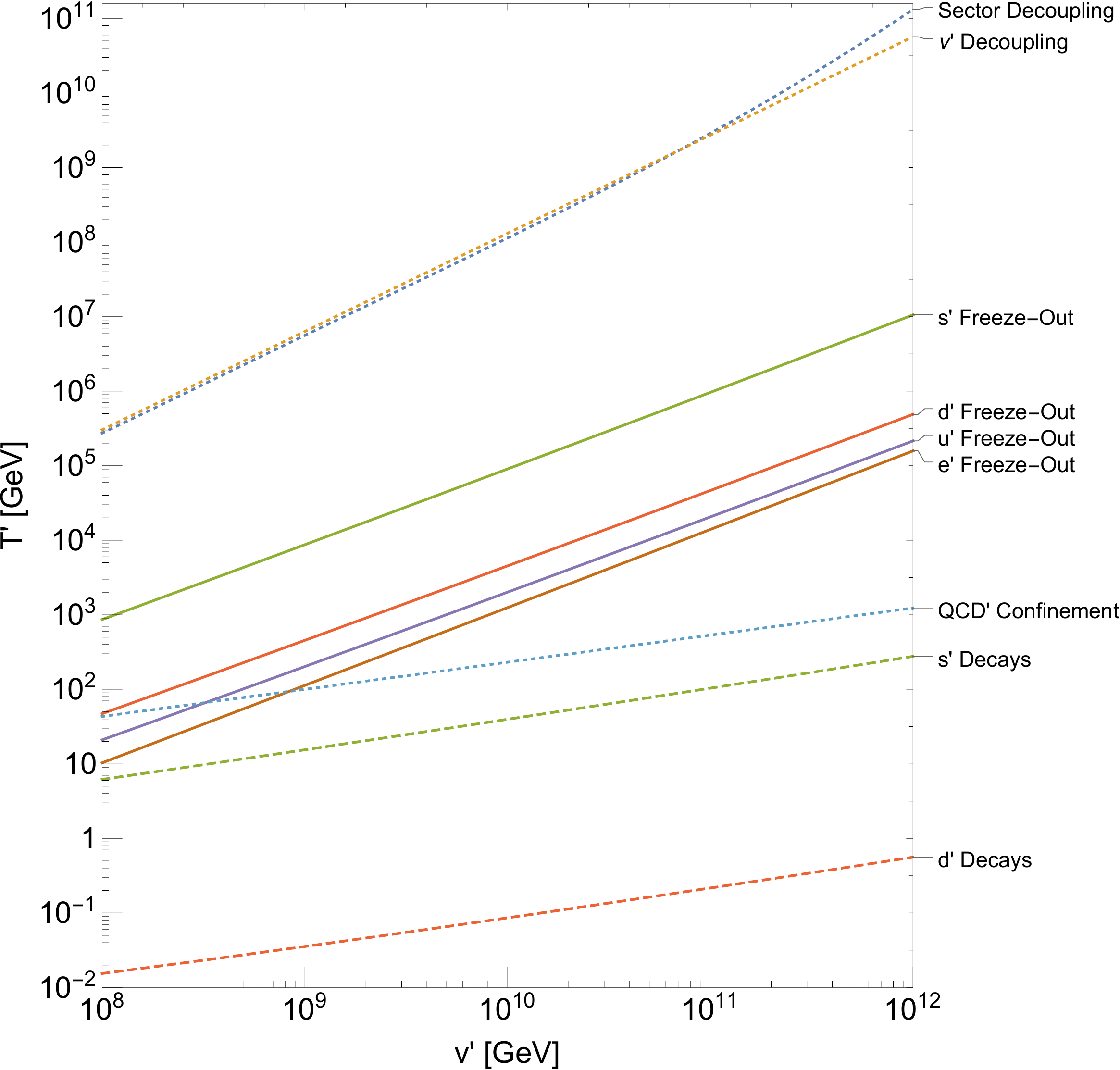} %
\caption{\small Temperatures of the mirror bath around which each mirror fermion freezes-out (solid) and decays (dashed). Mirror temperatures of sector decoupling, $\nu'$ decoupling, as well as the mirror QCD phase transition, are shown as dotted lines. }
\label{fig:tempFOplot}
\end{figure}
In this section, we take the reheat temperature of the universe larger than the temperature at which the two sectors decouple, $T_{\rm{RH}} > T_{\rm dec}$. In this case, the relic abundances of mirror $e'$ and $u'$ dark matter are set by freeze-out followed by dilution from the late decays of $\nu'$.\footnote{Furthermore, the maximum temperature of the universe after inflation is taken less than the mirror electroweak scale, to avoid domain wall problems from the spontaneous breaking of Higgs Parity. Generically the maximal temperature is higher than the reheat temperature. See~\cite{Harigaya:2013vwa,Mukaida:2015ria} for a recent estimation of the maximal temperature.}   As the temperature of the universe drops, unstable mirror particles decay, while stable $e'$ and $u'$ annihilate and freeze-out. Although heavier mirror charged leptons and quarks are unstable, their decay widths are much smaller than their masses because of the large mirror electroweak scale. Fig.~\ref{fig:tempFOplot} shows the temperatures around which each particle freezes-out (solid lines) and decays (dashed lines). Here we ignore the effects caused by late decays of mirror neutrinos, and include them momentarily.
For $v'$ in the range of $(10^8-10^{11})$ GeV, the $e'$ and $u'$ abundances are determined by the following processes in chronological order:
\begin{enumerate}
\item
$b'$ freezes-out.
\item
$c'$, $\mu'$ and $s'$ freeze-out. During these annihilations, $b'$ and $c'$ decay producing $c'$, $\mu'$ and $s'$. The annihilations also produce $e'$, $u'$ and $d'$, but they thermalize quickly.
\item
$d'$, $u'$ and $e'$ freeze-out. During these annihilations, $s'$ and $\mu'$ partially decay producing $e'$, $u'$ and $d'$.
\item QCD' phase transition occurs. Mirror hadrons composed of $s'$, $u'$ and $d'$ quickly annihilate. Mirror hadrons composed of $s'$ and $d'$ decay into $u'u'u'$.
\end{enumerate}
We note that $\tau'$ is short-lived and does not affect the above processes. A set of Boltzmann equations describing the freeze-out dynamics is shown in Appendix~\ref{sec:boltzmann}.

We elaborate on the fourth process. After the mirror QCD phase transition, mirror quarks are tied with each other by strings and form bound states. For $v' < 10^{10} ~\GeV$, the Coulomb binding energy of mirror hadrons containing a $u'$ or $d'$ is comparable to $T'_c$~\cite{Harigaya:2016nlg}, and so an $O(1)$ fraction of these mirror quarks form loosely bound states with large radii ~$\sim \Lambda_{\rm QCD}'$. With such a large cross-section, these mirror hadrons scatter among themselves efficiently, rearranging their quark constituent until they contain a $q' \bar{q}'$ pair,  and subsequently annihilate into $\gamma'$ ~\cite{Kang:2006yd,Harigaya:2016nlg}. For $v' > 10^{10} ~\GeV$ the Coulomb binding energy of mirror hadrons is larger than $T'_c$, and so most of the mirror quarks initially form tightly bound states with a smaller radius $\sim (m_{q'} \alpha_S')^{-1}$~\cite{DeLuca:2018mzn}. Nevertheless, these tightly bound states still have a relatively large radius and scatter and annihilate relatively efficiently. The mirror baryon containing only mirror strange quarks, $s's's'$, generally forms a tightly bound state for all $v'$. Still, $s'$ annihilates efficiently so that its beta decay contributions to $e'$ are small.  %

\begin{figure}[tb]
\centering
\includegraphics[width=0.8\textwidth]{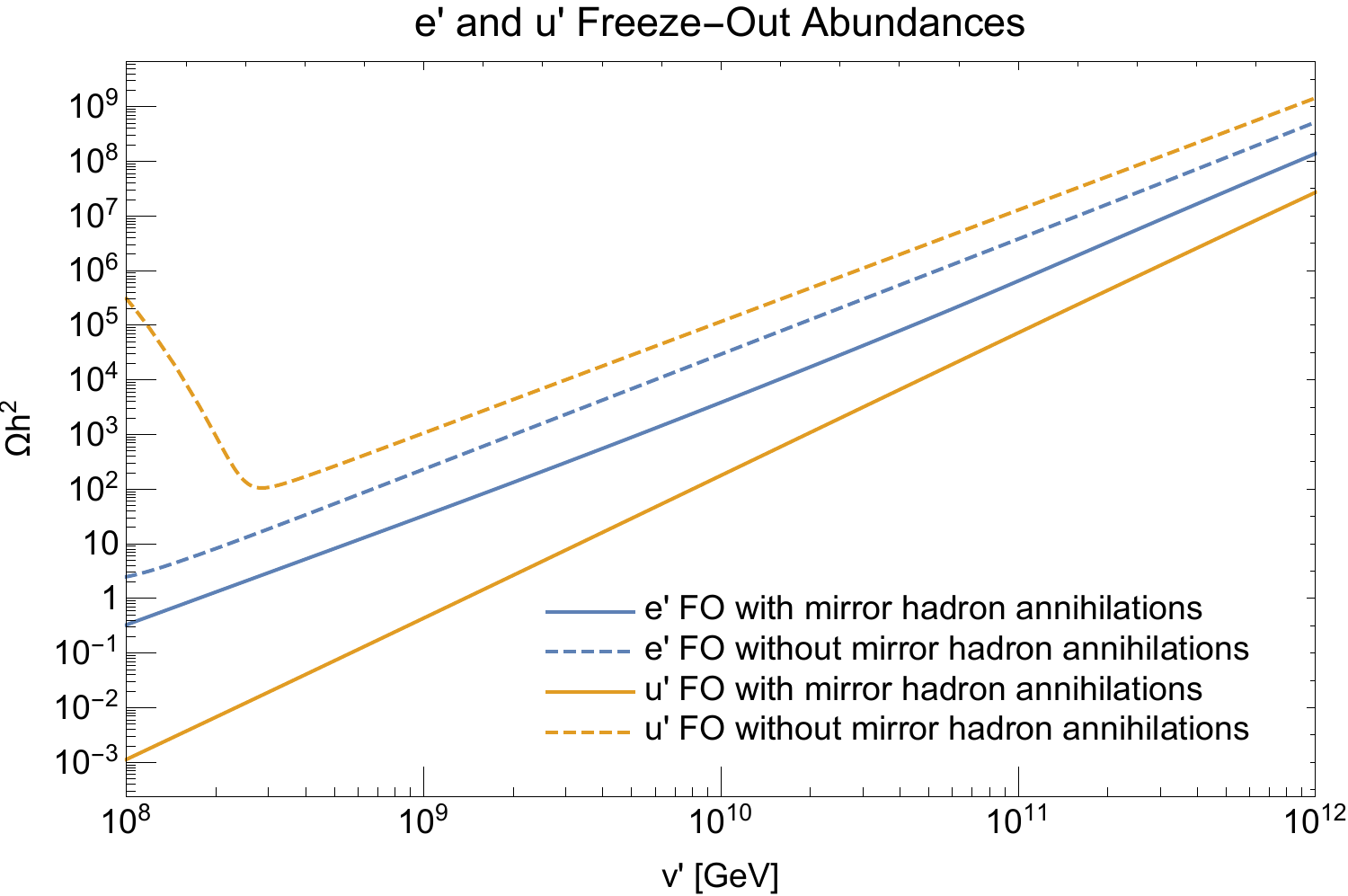} %
\caption{\small The cosmological abundance of mirror electrons and up quarks from freeze-out and from decays of heavier charged mirror fermions.  Dilution from mirror neutrino decays is not included.}
\label{fig:abundances}
\end{figure}

The thermal abundances of $e'$ and $u'$ are shown in Fig.~\ref{fig:abundances}. The solid lines conservatively assume that the annihilation cross-section of mirror hadrons is $\pi / (m_{q'} \alpha_S')^2$. The abundance of $e'$ does not change even if the cross-section is as large as  $ \Lambda_{\rm QCD}'^{-2}$. For comparison, the dashed line assumes mirror hadrons completely cease annihilating after confinement. Even though the annihilation cross-section of $e'$ does not change in either case, the relic abundance of $e'$ drops when annihilations of mirror hadrons continue after the QCD' phase transition since any beta decays from $s'$ or $d'$ that produce $e'$ below $T'_c$ are effectively absent (see Fig.~\ref{fig:tempFOplot}). To the left of the vertical dotted line, the QCD' phase transition occurs before $u'$ freezes-out, which is why its abundance dramatically increases if hadronic annihilations are assumed to cease below $T'_c$.

We see from the solid lines of Fig.~\ref{fig:abundances} that $e'$ is the dominant component of DM. On the other hand, efficient annihilations after the QCD$'$ phase transition make $u'$ a small component of DM, which exists today in the form of mirror hadrons like $u'u'u'$. For all $v' > 6 \times 10^7 ~\GeV$, the thermal abundance of $e'$ is too large to be DM. This is problematic as such a low $v'$ requires $m_t$ and $\alpha_S(m_Z)$ to lie beyond their current $3 \sigma$ experimental values. 

Nevertheless, in the above discussion, we have ignored mirror neutrinos which are cosmologically stable if  $m_{\nu'} < m_{e'} + m_{u'} + m_{d'}$ and $M_D$ of (\ref{eq:L}) is sufficiently large.  The former prevents decays to the mirror sector, due to mirror fermion number and mirror electromagnetic charge conservation, and the latter suppresses decays to the SM sector.  However, as $M_D$ is reduced, mirror neutrinos can decay well after they becoming non-relativistic to SM particles, thereby diluting $e'$ and $u'$. Consequently, the $v'$ required to produce $e'$ DM shifts to higher scales. 

Shortly after the two sectors decouple at $T_{\rm{dec}}$, $\nu'$ decouple from the mirror thermal bath as the mirror weak interaction rate drops below the Hubble expansion rate, as shown in Fig.~\ref{fig:tempFOplot}. Since $T_{\rm{dec}} = T'_{\rm{dec}} \gg m_{\nu'}$, $\nu'$ decouple while relativistic with an initial yield $Y_{\nu'} \simeq n_{\nu'}(T_{\rm{dec}})/s(T_{\rm{dec}})  = 0.0123$.  With this initial abundance, if $\nu'$ are sufficiently long-lived they dominate the energy density of the universe prior to decaying. 

\subsubsection{One generation of long-lived $\nu'$}
For our first example, we assume that two flavors of $\nu'$ decay rapidly and study $e'$ dilution from decays of the single long-lived flavor.   The long-lived $\nu'$ decays to $\ell H$ via the neutrino portal operator of (\ref{eq:L})\footnote{We take $\xi = \eta =1$.} with a decay rate
\begin{align}
\Gamma_{\nu' \to l h} = \frac{m_{\nu'}}{8 \pi}\frac{v'^2}{M_D^2}.
\label{eq:nulhdecay}
\end{align}
The mass of the mirror neutrino is given by Eq.~(\ref{eq:mnu}), and for sufficiently large $v'$, the mirror neutrino is massive enough that it can beta decay into $e', u'$ and $\bar{d}'$, with a decay rate
\begin{align}
\Gamma_{\nu' \to e'u'\bar{d}'} = \frac{3}{8}\frac{1}{192 \pi^3}\frac{m_{\nu'}^5}{v'^4}.
\end{align}
When $\nu'$ dominantly decay into the SM sector, the decay products heat up the SM thermal bath,  thereby diluting the frozen-out abundance of $e'$ and $u'$ relative to $n_\gamma$ by a factor
\begin{align}
    D = \frac{T_{\rm{MD},\nu'}}{T_{\rm{RH},\nu'}} \simeq \frac{m_{\nu'}Y_{\nu'}}{1.2 (\Gamma_{\nu'}M_{\rm{Pl}})^{1/2}}\left(\frac{\pi^2}{10g_{*RH}}\right)^{1/4}.
    \label{eq:dil}
\end{align}
($\Gamma_{\nu'})^{-1} = (\Gamma_{\nu' \to l h} + \Gamma_{\nu' \to e'u'\bar{d}'})^{-1}$ is the lifetime of the mirror neutrino. The numerical factor of $1.2$ is taken from~\cite{Harigaya:2018ooc}.
We solve the Boltzmann equation for the abundance of mirror fermions in Appendix~\ref{sec:boltzmann}, including freeze-out, the change of the expansion rate during the mirror neutrino matter-dominated era, and dilution from $\nu'$ decays. An approximation for the resulting $e'$ yield from freeze-out and dilution is
\begin{align}
	\frac{\rho_{e',\rm{FO}}}{s} \approx 35 \, \frac{m_{e'}^2}{\pi \alpha^2}\frac{1}{M_{\rm{Pl}}} \frac{g_*^{1/2}}{g_{*S}} \frac{1}{D} \approx 5 \times 10^{-6} \frac{v'^2 v}{\sqrt{M_{\rm{Pl}} m_\nu}}\frac{1}{M_D}
	\label{eq:FOest}
\end{align}
where $D$ is the dilution factor provided by mirror neutrino decays \eqref{eq:dil}.
\begin{figure}[tb]
\centering
\includegraphics[width=0.7\textwidth]{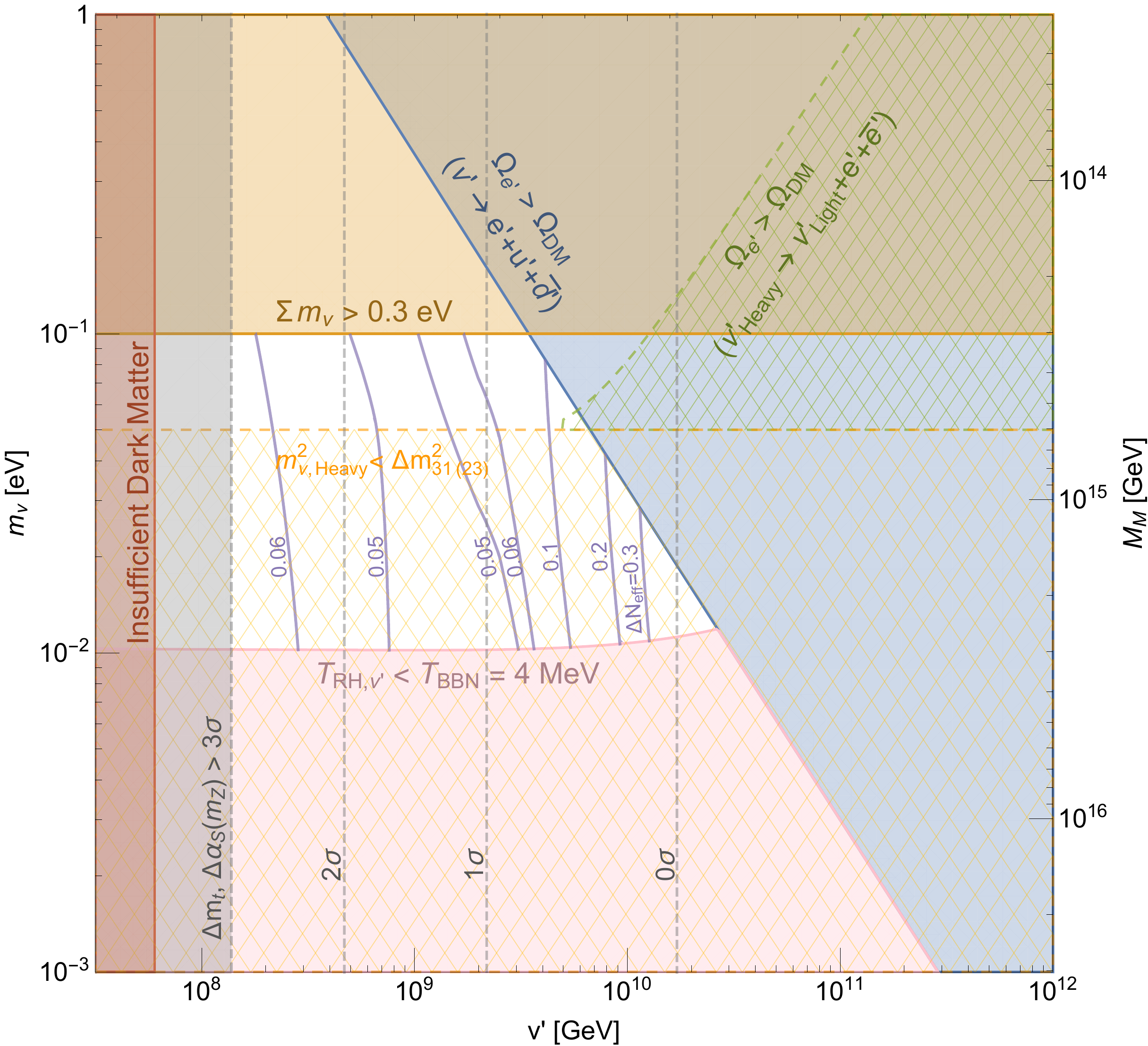} %
\caption{\small Constraints on $(v', m_{\nu})$ when $e'$ dark matter arises from freeze-out and dilution from one long-lived species of $\nu'$.  Here $m_\nu$ is the mass of the neutrino that is the Higgs Parity partner of the long-lived $\nu'$.  Purple contours show $\Delta N_{\rm eff}$ resulting from decays of $S'$ to $\gamma'$.
Vertical gray contours show $v'$ when $m_t$ and $\alpha_S(m_Z)$ deviate from their central values by $0$ to $3\sigma$.}
\label{fig:vp-mnu}
\end{figure}

For a given $(v', m_{\nu}')$, the parameter $M_D$ is determined to yield the correct $e'$ DM abundance. Furthermore,  the resulting values of $M_D$ are large enough that $m_{\nu}'$ can be mapped to $m_\nu$ by the scaling 
\begin{align}
    m_\nu = m_{\nu'} \frac{v^2}{v'^2}.
\end{align}
Further constraints on this scenario are shown in the $(v', m_{\nu})$ plane in Fig.~\ref{fig:vp-mnu}.

In the allowed white region, we find $M_D$ must lie within the range $(10^{18} - 10^{23}) ~\GeV$. In the red-shaded region, the $e'$ abundance is smaller than the dark matter abundance without dilution. 
For too small a neutrino mass, the required $T_{\rm{RH},\nu'} \approx \sqrt{\Gamma_{\nu'}M_{\rm{Pl}}}$ to reproduce the dark matter abundance is below the MeV scale and affects BBN as well as the effective number of neutrinos~\cite{Kawasaki:1999na,Kawasaki:2000en}. We adopt the bound $T_{\rm{RH},\nu'} > 4$ MeV~\cite{deSalas:2015glj}, excluding the pink-shaded region.
In the blue-shaded region, the mirror beta decay $\nu'\rightarrow e' u'\bar{d}'$ is kinematically allowed, creating too much $e'$ and $u'$ abundance.  In the orange-shaded region the sum of the SM neutrino masses are above 0.3 eV, which is disfavored by the observations of the Cosmic Microwave Background (CMB)~\cite{Aghanim:2018eyx}.  The gray-shaded region is excluded at the $3\sigma$ level from measurement of $\alpha_S$ and the Higgs and top masses. If the long-lived species is the lightest $\nu'$ then beta decay to $\nu' e'\bar{e}'$ cannot occur.  However, if the long-lived $\nu'$ is one of the heavier states, then the lightly green-shaded region of Fig.~\ref{fig:vp-mnu} is also excluded since the long-lived $\nu'$ creates $e'$. The corresponding SM neutrino mass should be above $\Delta m^2_{31(23)}$, excluding the lightly yellow-shaded region.  The allowed region is not large: $m_t$ should be above its present central value and, remarkably, the neutrino mass must be within a factor of 10 of its present upper bound of $0.1$ eV. 

In the resulting allowed region of parameter space for $e'$ dark matter, the purple contours show our prediction for $\Delta N_{\rm{eff}}$ from decays of mirror glueballs, produced at the QCD$'$ confining transition, to mirror photons.  Throughout the entire region $\Delta N_{\rm{eff}}$ is in the range $0.03 \mathchar`- 0.4$, allowed by Planck~\cite{Aghanim:2018eyx} and within range of the sensitivities of CMB Stage IV experiments~\cite{Abazajian:2016yjj}.

\subsubsection{Universal coupling strength of neutrino portal}

As a second illustration of $e'$ freeze-out and dilution from $\nu'$ decays, we take the strength of the neutrino portal coupling to be independent of generation.  Thus, in a neutrino mass basis, we take $\nu' \rightarrow lH$ decays to be given by (\ref{eq:nulhdecay}) for all three generations of $\nu'$.   To avoid overproducing $e'$,  all three $\nu'$ must be light enough that beta decay is forbidden. Thus the total decay rate of each mirror neutrino is given by \eqref{eq:nulhdecay} and is proportional to $m_{\nu'}$. Consequently, the dilution \eqref{eq:dil} is dominated by the heaviest mirror neutrino. For a normal hierarchy ($m_{\nu_1} \ll m_{\nu_2} < m_{\nu_3}$) of SM neutrinos, the mirror neutrino responsible for dilution is $\nu'_3$; for an inverted hierarchy ($m_{\nu_3} \ll m_{\nu_1}<m_{\nu_2}$), $\nu'_{2,1}$ give comparable dilutions; and for a quasi-degenerate spectrum $\nu'_{3,2,1}$ all give comparable dilutions. 

The bounds from BBN, too much dark matter from $\nu'\rightarrow e' u'\bar{d}'$ decay, and too little dark matter from freeze-out are approximately as in Fig.~\ref{fig:vp-mnu}, with the vertical axis interpreted as the heaviest neutrino, which is constrained by oscillation data to be at or above 0.05 eV.  Thus the larger values of $v'$ and $\Delta N_{\rm{eff}}$ are excluded in this case.  The upper bound on the heaviest neutrino from the cosmological limit on the sum of the neutrino masses is 0.1 eV.
\begin{figure}[tb]
\centering
\includegraphics[width=0.7\textwidth]{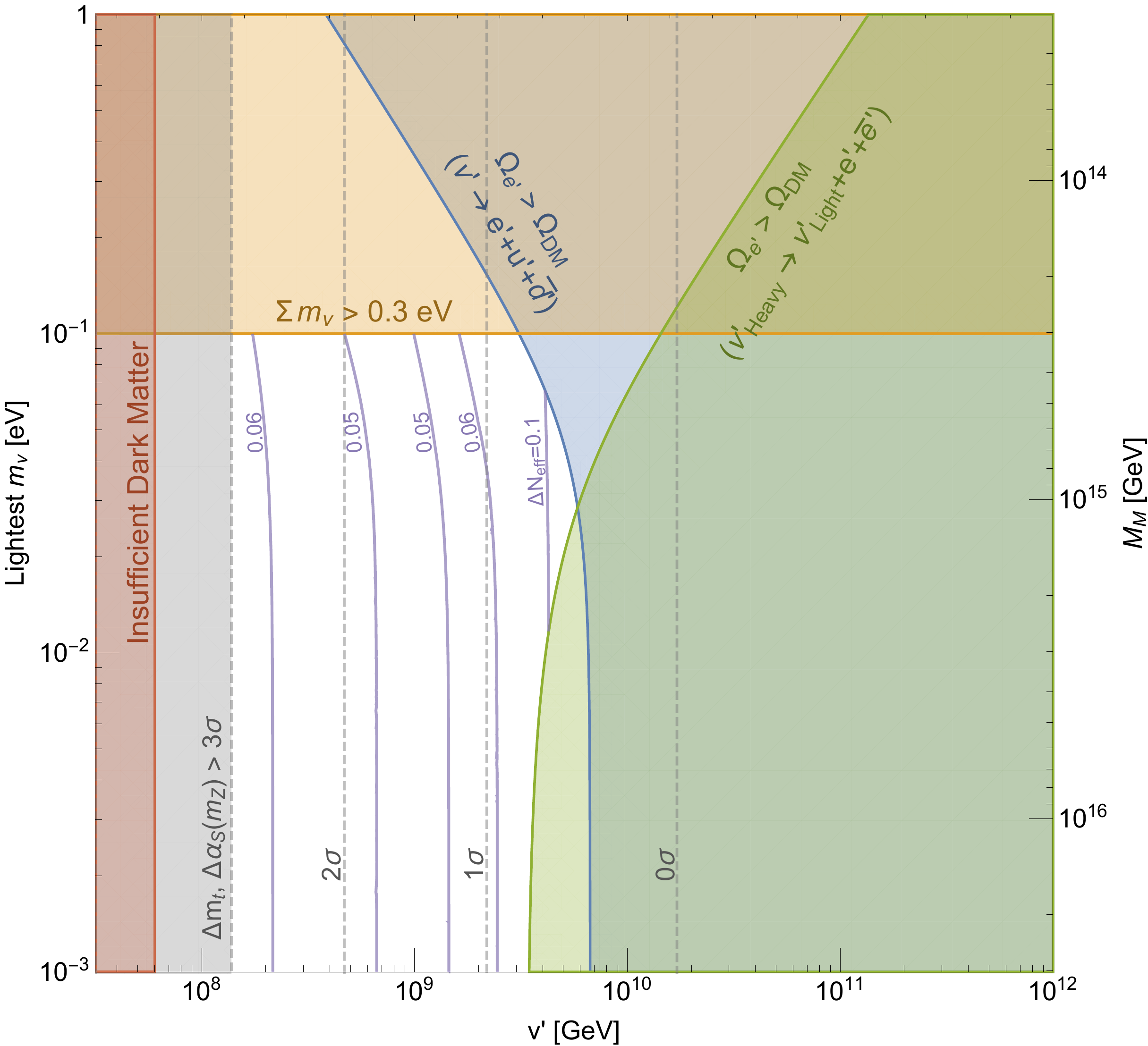} %
\caption{\small Constraints on $(v', m_{\nu})$ when $e'$ dark matter arises from freeze-out and dilution from $\nu'$ with universal neutrino portal couplings.  Here $m_\nu$ is the mass of the {\it lightest} neutrino.  Purple contours show $\Delta N_{\rm eff}$ resulting from decays of $S'$ to $\gamma'$.
Vertical gray contours show $v'$ when $m_t$ and $\alpha_S(m_Z)$ deviate from their central values by $0$ to $3\sigma$. In the allowed white region, $\Delta N_{\rm{eff}}$ is always greater than $0.03$, which will be probed by CMB Stage IV~\cite{Abazajian:2016yjj}.}
\label{fig:vp-mnu3}
\end{figure}

In addition to these bounds, there is a constraint from the decay $\nu_3' \rightarrow \nu_{1,2}' e'\bar{e}'$ for a normal hierarchy or $\nu_{2,1}' \rightarrow \nu_{3}' e'\bar{e}'$ for an inverted hierarchy. In either case, too much $e'$ is produced. Regardless of whether the SM neutrinos obey a normal or inverted hierarchy, this constraint can be translated to a bound on the \textit{lightest} SM neutrino:
\begin{align}
    m_{\nu,{\rm lightest}} > \frac{\Delta m_{31}^2}{4 m_e} \frac{v'}{v} - m_e \frac{v}{v'} .
    \label{eq:nu'cascade}
\end{align}
$\Delta m_{31}^2 \equiv |m_3^2 - m_1^2| \simeq (0.05 ~{\rm eV})^2$ is the atmospheric neutrino mass difference squared and $m_e$ is the electron mass. We have made the good approximation that $\Delta m_{31}^2$ is also the mass squared difference between the lightest and heaviest SM neutrino in an inverted hierarchy. This bound is shown in the yellow hatched region of Fig.~\ref{fig:vp-mnu}.

The constraints on this scheme for $e'$ dark matter are shown in Fig.~\ref{fig:vp-mnu3}, where the vertical axis is the {\it lightest} SM neutrino mass.  The bound of \eqref{eq:nu'cascade} appears in green. If $v'$ turns out to be larger than $4\times 10^9$ GeV, the lightest neutrino mass is predicted to be in a narrow range.  The lightest mirror neutrino is longer-lived than the heaviest mirror neutrino for a universal $M_D$, but decays before the onset of the BBN for $m_\nu > 10^{-3}$ eV.

The sum of the masses of the three neutrinos can be constrained through its imprint on the  structure of the universe.
Future measurements of the CMB, BAO, and 21 cm emission are expected to determine the sum of the masses with an uncertainty of $10$ meV ~\cite{Font-Ribera:2013rwa,Allison:2015qca,Archidiacono:2016lnv}.
One can check the consistency of the the measurements and the bounds we have obtained.

During the matter dominated era by $\nu'$, cosmic perturbations of massive components can grow. Since $e'$ tightly couples to mirror photons, the perturbation of $e'$ does not grow by itself. The perturbation of mirror glueballs grows, decays into mirror photons, which scatter with $e'$ and grow the perturbation of $e'$, like the growth of a weakly interacting massive particle during a matter dominated era~\cite{Choi:2015yma}. We will discuss the implication of the growth to the future searches for ultra compact mini halo elsewhere.

\subsection{Freeze-In from Higgs Portal and Kinetic Mixing}
\label{sec:fi}
\begin{figure}[tb]
    \centering
    \begin{minipage}{0.455\textwidth}
        \centering
        \includegraphics[width=.675\textwidth]{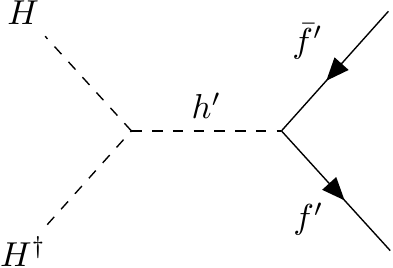} 
    \end{minipage}\hfill
    \begin{minipage}{0.455\textwidth}
        \centering
        \includegraphics[width=1\textwidth]{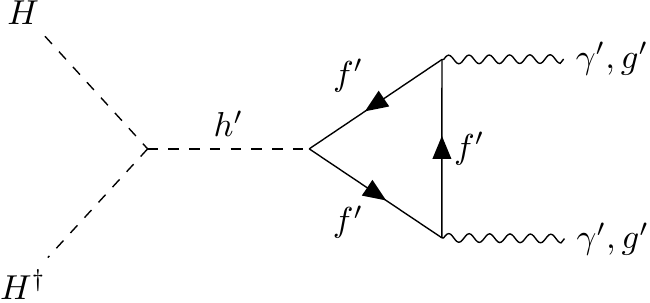} 
    \end{minipage}
    \caption{\small Freeze-in production of mirror fermions (left) and mirror gauge bosons (right) through the Higgs portal.}
    \label{fig:fpHiggsPortal}
\end{figure}

In this section, we consider the relic abundances of mirror $e'$ when the reheat temperature of the universe is below $T_{\rm dec}$ and only the SM sector is reheated. Since the SM and mirror sectors are weakly coupled below $T_{\rm{dec}}$, mirror DM is produced via freeze-in through the Higgs portal, as shown in Fig.~\ref{fig:fpHiggsPortal}.
Although the mirror fermion and gauge boson production rates are UV-dominated, the entropy production during reheating negates far-UV production so that the dominant production occurs around $T_{\rm{RH}}$. Reheat temperatures below the mirror electron mass yield insufficient $e'$ to be DM since the small $e'$ freeze-in abundance is further diluted by $(m_{e'}/T_{\rm{RH}})^6$ as production almost ceases below $T \approx m_{e'}$.%
\footnote{Some $e'$ production still occurs for $T_{\rm{RH}} < T <  m_{e'}$ by scatterings involving highly energetic particles produced by inflatons~\cite{Harigaya:2014waa,Harigaya:2019tzu},
which we find is not efficient enough to reproduce the DM abundance.}
Consequently, we focus on $T_{\rm{RH}} \gtrsim m_{e'}$. A set of Boltzmann equations describing the freeze-in dynamics is shown in Appendix~\ref{sec:boltzmann}. The thermal evolution of
the mirror electrons is as follows:

At $T_{\rm{RH}}$, the mirror electrons carry a typical energy $T_{\rm{RH}}$ and a freeze-in number density%
\footnote{For low $v'$ and high $T_{\rm{RH}}$, $e'$ and $\gamma'$ may thermalize during reheating, altering (\ref{eq:nFI}). Thermalization cools the mirror bath so that mirror particles freeze-out instantly but are then replenished by the Higgs portal. Since freeze-in production is maximized at $T_{\rm{RH}}$, any pre-thermalized contribution is typically small. Even so, we consider this effect in Appendix \ref{sec:boltzmann}.}
\begin{align}
	n(T_{\rm{RH}}) \, = \, \frac{4}{9} \frac{n_H(T_{\rm{RH}})^2}{H(T_{\rm{RH}})} \; \langle \sigma v  (T_{\rm{RH}})\rangle.
\label{eq:nFI}
\end{align}
$n_H$ is the SM Higgs thermal number density, $H$ is Hubble, and $\langle \sigma v\rangle$ is the freeze-in cross-section given by
\begin{align}
\langle \sigma v(T_{\rm{RH}})\rangle = \frac{1}{8 \pi} \frac{y_e^2}{v'^2}.
\end{align}

\begin{figure}[tb]
\centering
\includegraphics[width=0.6\textwidth]{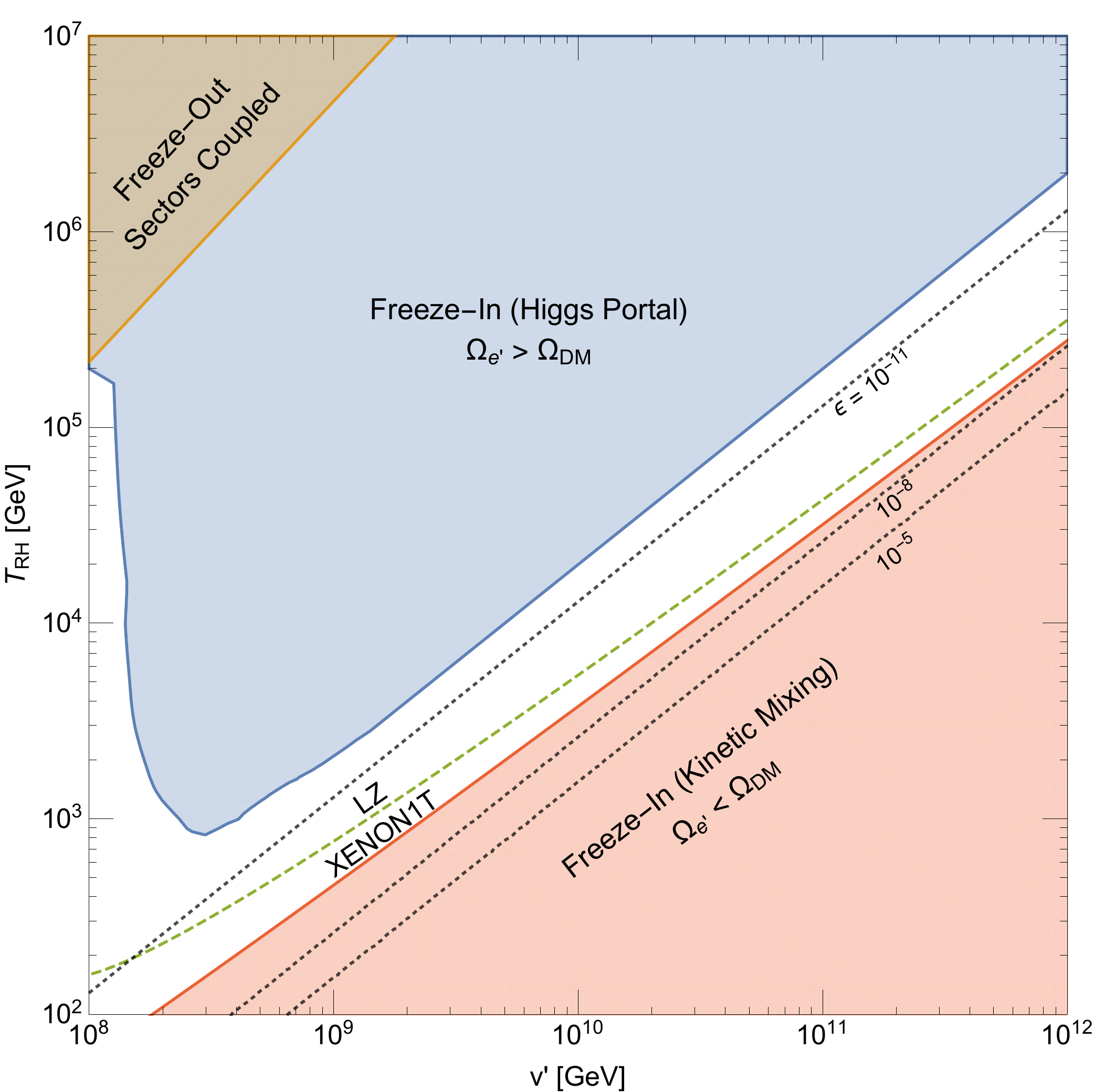} %
\caption{\small Constraints on the mirror electroweak scale $v'$ and the reheat temperature $T_{\rm{RH}}$ of the universe. In the blue region, $e'$ is overproduced via freeze-in from the Higgs portal. In the red region, the required $\epsilon$ to freeze-in $e'$ as DM via the kinetic mixing portal (shown by the dotted counters) is large enough to produce nuclear recoil signals in XENON1T. In the orange region, the reheat temperature is high enough that the two sectors were originally thermally coupled and the freeze-in regime reduces to the freeze-out regime (see Sec. \ref{sec:fo}).}
\label{fig:vp-trh}
\end{figure}

For all $v'$, the frozen-in abundance of $e'$ at $T_{\rm{RH}}$ exceeds that of dark matter for $T_{\rm{RH}} \gtrsim m_{e'}$. For $v' \gtrsim 4 \times 10^8 ~\GeV$, annihilations of $e'$ are ineffective during subsequent freeze-out. The freeze-in yield of $e'$ from the Higgs portal is
\begin{align}
	\frac{\rho_{e',\rm{FI}}}{s} &\approx 0.01 \frac{1}{(g_*)^{1/2} g_{*S}} \frac{y_e^3}{v'}T_{\rm{RH}} M_{\rm{Pl}} &\text{(Higgs Portal)}
	\label{eq:hPortalFI}
\end{align}
In this regime, a reheat temperature approximately equal to the mirror electron mass reproduces the correct DM abundance, as shown in Fig.~\ref{fig:vp-trh}.

For $v' \lesssim 4 \times 10^8 ~\GeV$, annihilations of $e'$ are effective during subsequent freeze-out and the allowed $T_{\rm{RH}}$ rises, as shown in Fig.~\ref{fig:vp-trh}. However, as $T_{\rm{RH}}$ increases, mirror fermions heavier than $e'$ are produced at $T_{\rm{RH}}$, which transfer much of their abundance to $\gamma'$ and $e'$ as they annihilate and thermalize via $2 \to 2$ and $2 \to 3$ processes as discussed in Appendix~\ref{sec:boltzmann}.

For $T_{\rm{RH}} \geq T_{\rm{dec}}$, the two sectors were once in thermal equilibrium and the situation reverts to traditional freeze-out discussed in Sec. \ref{sec:fo}.  $\Delta N_{\rm eff}$ and BBN constraints from frozen-in mirror glueball decays are not shown in Fig.~\ref{fig:vp-trh} as they are much weaker than the bound on overproduction of $e'$ DM. 

Finally, as mentioned in Sec. \ref{sec:kineticMixing}, $e'$ DM can also be frozen-in via kinetic mixing induced from higher dimensional operators \eqref{eq:eps1}. On one hand, the freeze-in abundance of $e'$ through the Higgs portal is dominantly set by its yukawa coupling, which is fixed and whose smallness prevents sufficient $e'$ to be produced as DM for $T_{\rm{RH}} < m_{e'}$. On the other hand, the freeze-in abundance of $e'$ through kinetic mixing is set by $\epsilon$, which is a free parameter (indirectly set by the unification scale $v_{G}$), and whose value can be chosen to sufficiently produce $e'$ DM for reheat temperatures as low as $\sim m_{e'}/25$. 

For $T_{\rm{RH}} < m_{e'}$, the freeze-in yield of $e'$ from kinetic mixing is
\begin{align}
	\frac{\rho_{e',\rm{FI}}}{s} &\approx 0.02 \, \pi \alpha^2 \epsilon^2 M_{\rm{Pl}}\left(\frac{m_{e'}}{T_{\rm{RH}}}\right)^2 \exp\left(-\frac{2m_{e'}}{T_{\rm{RH}}}\right) &\text{(Kinetic Mixing)}
	\label{eq:kineticMixingFI}
\end{align}
The black dotted contours in the region $T_{\rm{RH}} < m_{e'}$ of Fig.~\ref{fig:vp-trh} show the $\epsilon$ necessary for $e'$ to be frozen-in as DM.
The shaded red region is excluded if $e'$ is the DM since the required $\epsilon$ to freeze-in $e'$ DM via kinetic mixing is large enough to already produce recoil signals at XENON1T\footnote{If $e'$ is not the DM, or is produced in a non-thermal way, the red region is not applicable and the $SM \times SM'$ model is not necessarily excluded.}. A similar calculation for the proposed LZ experiment, which can probe $\epsilon$ an order of magnitude smaller, produces the green contour `LZ'. For low $v'$, LZ has the potential to probe nearly all reheat temperatures capable of freezing-in $e'$. 
\section{Gravitational Waves from Mirror QCD phase transition}
\label{sec:GW}

In the range of $v'$ consistent with the observed top quark mass, mirror quark masses are much larger than the mirror QCD scale. The mirror QCD phase transition is then first order~\cite{Yaffe:1982qf,Svetitsky:1982gs}.
The phase transition proceeds by nucleation of bubbles, which collide with each other and produce gravitational waves~\cite{Witten:1984rs}.

We consider the case where the $e'$ dark matter abundance is set by freeze-out followed by dilution from late $\nu'$ decays. The abundance of gravitational waves $\Omega_{\rm GW,col} h^2$ directly produced by the bubble collisions as a function of a frequency $f$ is given by
\begin{align}
\frac{d \, \Omega_{\rm GW,col} \, h^2}{d \, {\rm ln} f} \; \simeq \; & 2 \times 10^{-8} \;  \frac{ (f / f_p)^3}{ 0.3 + (f/f_p)^4} \, \left( \frac{10}{\beta/H} \right)^2 D^{-4/3} \left( \frac{\rho_{g'}/\rho_{\rm tot}}{2/3}  \frac{\rho_{\rm lat}}{\rho_{g'}}   \frac{\rho_{\rm kin}}{\rho_{\rm lat}} \right)^2  \frac{\rho_{\rm tot} / \rho_{\rm SM}}{3}  , \label{eq:GW}\\
f_p \; \simeq \; & 2\times 10^{-5} \, {\rm Hz}  \left(\frac{\beta/H}{10} \right) \, \left( \frac{T'_c}{ 100\, {\rm GeV}} \right) D^{-1/3} \; \left( \frac{g_{\rm dec}'}{60} \frac{100}{g_{\rm dec}} \right)^{1/3}  \; \left( \frac{ \rho_{\rm tot} / \rho_{g'} } {3/2} \right)^{1/2} \left( \frac{b'}{0.5} \right)^{1/6}.
\end{align}
$f_p$ is close to the frequency at the peak of the distribution and $T'_c \simeq 1.3 \, \Lambda'_{\rm QCD}$ is the temperature of the mirror QCD phase transition.
Here we use the results of
Ref.~\cite{Huber:2008hg}, assuming that the velocity of the bubble wall is the speed of light, and take into account the dilution $D$ from $\nu'$ decay.
The ratio $(\beta/H)$ parametrizes the duration of the phase transition $\beta^{-1}$ in comparison with the Hubble time scale $H^{-1}$. $\rho_{\rm tot}$ is the total energy density, $\rho_{g'}$ is the energy density of the mirror gluon bath, $\rho_{\rm lat}$ is the latent heat of the phase transition, $\rho_{\rm kin}$ is the kinetic energy of the bubble wall and $\rho_{\rm SM}$ is the energy density of the SM bath, all of which are evaluated at the phase transition. $g_{\rm dec}$ and $g_{\rm dec}'$ are the degrees of freedom of the SM and the mirror sector at the decoupling of the two sectors, respectively. $b'$ parametrizes the energy density of the mirror gluon just before the phase transition, $\rho_{g'} = b' {T_{\rm QCD}'}^{4}$. 
The ratio $\rho_{\rm SM} / \rho_{g'}$ is estimated in Appendix~\ref{sec:rhos}.

\begin{figure}[tb]
\centering
\includegraphics[width=1.0\textwidth]{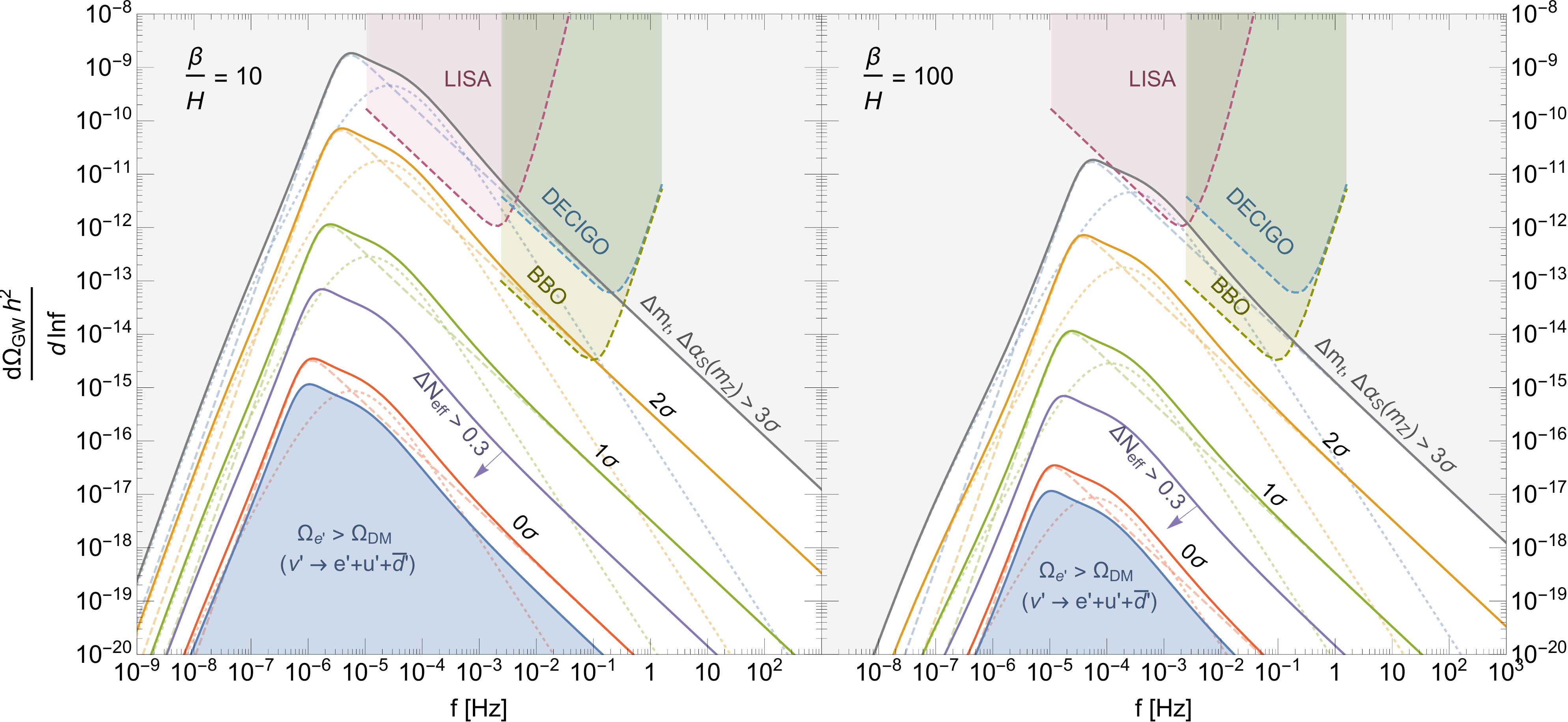} %
\caption{\small Gravitational wave spectrum generated by the mirror QCD phase transition for $\beta/H = 10$ (left) and $\beta/H = 100$ (right). Future gravitational wave detectors such as LISA and BBO may detect a signal if $m_t$ and $\alpha_S(m_Z)$ lie more than $2\sigma$ away from their current central values.}
\label{fig:GW}
\end{figure}

Gravitational waves are also produced by the turbulent motion of fluids induced by the bubbles~\cite{Kamionkowski:1993fg}. The abundance of such gravitational waves $\Omega_{\rm GW,tub} h^2$ is 
\begin{align}
\frac{d \, \Omega_{\rm GW,tub} \, h^2}{d \, {\rm ln} f} \; \simeq \; & 4 \times 10^{-9} \;   \frac{9 (f/f_p)^3}{( f / f_p + 0.02 H/\beta) ( f / f_p + 0.8 )^{11/3}} \left( \frac{10}{\beta/H} \right)^2 D^{-4/3} \nonumber \\
 & \times \left( \frac{\rho_{g'} / \rho_{\rm tot}}{2/3}  \frac{\rho_{\rm lat}}{\rho_{g'}}   \frac{\rho_{\rm kin}}{\rho_{\rm lat}} \right)^{3/2}  \frac{\rho_{\rm tot} / \rho_{\rm SM}}{3}
\label{eq:GW2} \\
f_p \; \simeq \; &  1\times 10^{-4}  {\rm Hz} \;\; \left(\frac{\beta/H}{10} \right) \, \left( \frac{T'_c}{ 100\, {\rm GeV}} \right) D^{-1/3} \; \left( \frac{g_{\rm dec}'}{60} \frac{100}{g_{\rm dec}} \right)^{1/3}  \; \left( \frac{ \rho_{\rm tot} / \rho_{g'} } {3/2} \right)^{1/2} \left( \frac{b'}{0.5} \right)^{1/6}.
\end{align}
Here we use the results of Refs.~\cite{Caprini:2009yp,Caprini:2010xv} assuming that the bubble walls expand at the speed of light.%
\footnote{Since the mirror QCD bath couples to the standard model particles very weakly, bubbles only induce turbulent motion of mirror glueballs. In particular, a turbulent magnetic field is not induced. For a phase transition generating magnetic turbulence, Ref.~\cite{Caprini:2009yp} finds a spectrum of gravitational waves produced by turbulent magnetic fields similar to that from turbulent motion of fluids, and hence we simply use the fitting provided  in Ref.~\cite{Caprini:2010xv}.}
Numerically, this contribution is smaller than the one from the bubble collision.

The prediction (\ref{eq:GW}, \ref{eq:GW2})  for the gravity wave spectrum depends on $v'$ via $T'_c$ and especially $D$.  With $v'$ determined by the top quark mass, we show in Fig.~\ref{fig:GW} the prediction for the spectrum of the gravitational waves for various $m_t$, taking $\beta/H$ of (10,100) in the (left, right) panel. 
The dashed and dotted lines show the contribution from the bubble collision and the turbulent motion respectively, and the solid lines show the sum of them.
In the blue shaded region, the freeze-out followed by the dilution from $\nu'$ fails as is shown in Fig.~\ref{fig:vp-mnu}.
The ratio $(\beta/H)$ is likely to be $O(100)$~\cite{Hogan:1984hx}.
If the top quark mass is large enough, gravitational waves can be detected by future experiments such as LISA, DECIGO and BBO~\cite{Moore:2014lga}. We note that prediction for the gravitational wave spectrum assumes that the phase transition occurs before the $\nu'$ matter-dominated era. This condition is satisifed in the region where future experiments may detect the gravitational wave spectrum, that is, at the $2-3 \, \sigma$ level for $m_t$ and $\alpha_S(m_Z)$.

We also 
note that many aspects of the phase transition in QCD-like theories, such as~($\beta/H$) and $\rho_{\rm kin} / \rho_{\rm lat}$, are not well-understood because of the non-perturbative nature. Once the phase transition is well-understood, it will become possible to check the consistency of future measurements of the top quark mass and the gravitational wave spectrum.

\section{Conclusions and Discussions}

We have introduced the Mirror Higgs Parity theory, described by (\ref{eq:L}).  The entire SM Lagrangian, including dimension 5 operators for neutrino masses,  is replicated by Higgs parity and the only unknown parameters are those of the kinetic mixing, Higgs and neutrino portals that connect the two sectors.  The spectrum of the mirror sector is a scaled up version of the SM spectrum, as shown for the light mirror particles in Fig.~\ref{fig:mirrorSpectrum}.  The scaling depends only on the Higgs Parity breaking scale $v'$, which sets the scale at which the SM Higgs quartic vanishes and will become better determined by precision measurements of $(m_t, \alpha_S)$.

There are several interesting theories containing the Higgs Parity mechanism for the vanishing of the Higgs quartic at high energies.   Mirror Higgs Parity is the simplest theory where the Higgs Parity partner of the electron, $e'$, is dark matter, with an abundance set by thermal mechanisms.  Direct detection of $e'$ dark matter can occur via kinetic mixing and leads to a recoil spectrum characteristic of photon exchange. The present bound from XENON1T and the future reach of LZ on the kinetic mixing parameter $\epsilon$ are shown in Fig.~\ref{fig:epsilonConstraintPlot}.

If the SM gauge group is unified at scale $v_G$ into a group such as SU(5), the proton decay rate scales as $\Gamma_p \propto 1/v_G^4$.  Furthermore, since kinetic mixing vanishes in the unified theory, it may arise from a higher dimensional operators, such as in Eq.~(\ref{eq:GUTkineticmixing}), leading to $\epsilon \propto v_G^n$, where $n$ is a model-dependent, positive integer.   Thus proton decay excludes small $v_G$ and direct detection excludes large $v_G$.  The correlation of these two rates for $n=4$ is shown in Fig.~\ref{fig:ddProtonDecay}.  A large fraction of the allowed parameter space of the theory will be probed by a combination of Hyper-Kamiokande and LZ.

For large values of the reheat temperature after inflation, $T_{\rm{RH}}$, the SM and mirror sectors reach thermal equilibrium via the Higgs portal interaction.  The $e'$ relic abundance arises first from freeze-out and is then diluted by $\nu'$ decay to $\ell H$.  Fixing the neutrino portal parameters to obtain the observed abundance, the remaining relevant parameters are $v'$, which determines $m_{e'}$, and $m_\nu$ which determine $m_{\nu'}$.   The constraints on this scheme for dark matter are shown in the $(v', m_\nu)$ plane in Fig.~\ref{fig:vp-mnu}, for the case that dilution is dominated by a single $\nu'$.  Remarkably,  the corresponding neutrino is required to have a mass larger than 0.01 eV, in the range of masses determined from oscillation data.  Furthermore, $v'$ must be in the range of $(10^8 - 10^{10})$ GeV, overlapping the allowed range determined by requiring the Higgs quartic to vanish at $v'$.  Within this allowed unshaded region of Fig.~\ref{fig:vp-mnu}, we predict the contribution to dark radiation arising from decays of mirror glueballs to mirror photons. The resulting $\Delta N_{\rm eff}$, shown by purple contours, varies from about 0.04 to 0.4, and is highly correlated with $v'$ and therefore with $m_t$.

Since all the mirror quarks are much heavier than the mirror confining scale, the mirror QCD phase transition, which occurs at $T' \sim (40-1000)$ GeV for $v' = (10^8 - 10^{12})$ GeV, is first order and produces gravitational waves from bubble dynamics and turbulent fluid motion at the transition. The spectral energy density today, normalized to the critical energy density, is then obtained by including the $\nu'$ decay dilution factor, and is shown in Fig.~\ref{fig:GW}.
Part of the allowed region of the theory can be probed by LISA, DECIGO and BBO, and a gravity wave signal in these experiments would be correlated with $m_t$ and $\Delta N_{\rm eff}$. 

For low values of the reheat temperature after inflation, $T_{\rm{RH}}$, $e'$ DM can arise via freeze-in production.  The observed DM abundance may be obtained anywhere in the unshaded region of Fig.~\ref{fig:vp-trh}.  On the edge of the blue shaded region this occurs via the Higgs portal, which is UV dominated around $T_{\rm{RH}}$.  In the rest of the unshaded region this occurs via kinetic mixing, dominated at temperatures near $m_{e'}$, for a suitable value of $\epsilon$.

Mirror Higgs Parity exchanges $SU(3)$ with $SU(3)'$ and hence does not solve the strong CP problem. One possible solution is to introduce a QCD axion~\cite{Peccei:1977hh,Peccei:1977ur,Weinberg:1977ma,Wilczek:1977pj}. If Higgs Parity transforms the QCD axion into a mirror QCD axion, the mirror QCD axion is an axion-like-particle with a mass 
\begin{align}
m_{a'} = 0.6 \frac{\Lambda_{\rm QCD}^{'2}}{f_a} =  0.4~{\rm keV} \left( \frac{v'}{10^9~{\rm GeV}} \right)^{8/11} \frac{10^{10}~{\rm GeV}}{f_a},
\label{eq:ma'}
\end{align}
where the topological susceptibility is taken from~\cite{Durr:2006ky}.
The mass is correlated with $v'$ and hence with the top quark mass. Both axions may contribute to the dark matter density.

Alternatively, if the QCD axion is neutral under Higgs Parity it couples to QCD and mirror QCD with the same decay constant. Since Higgs Parity ensures the equality of the theta angles in the two sectors, the strong CP problem is still solved~\cite{Rubakov:1997vp,Berezhiani:2000gh,Hook:2014cda,Fukuda:2015ana}. The mass is given by Eq.~(\ref{eq:ma'}). An advantage of such a heavy axion is that it is easier to understand the PQ symmetry as an accidental symmetry~\cite{Fukuda:2015ana}. In this case, it is even possible to have a small decay constant $\ll 10^9$ GeV, since the large mass prevents the production of axions in stellar objects and meson decays.  We will discuss the phenomenology of axion dark matter in Mirror Higgs Parity in future works.

\section*{Acknowledgement}
This work was supported in part by the Director, Office of Science, Office of High Energy and Nuclear Physics, of the US Department of Energy under Contracts DE-AC02-05CH11231 and DE-SC0009988 (KH), as well as by the National Science Foundation under grants PHY-1316783 and PHY-1521446. 
This work was performed in part at Aspen Center for Physics, which is supported by National Science Foundation grant PHY-1607611.


\appendix

\section{Boltzmann equations for the $e'$ and $u'$ abundance}
\label{sec:boltzmann}
In this appendix we show the Boltzmann equations governing the thermal relic abundance of  $e'$ and $u'$. To simplify the expression, we omit the superscript $'$ except for the titles of sections and the mirror temperature $T'$. The number densities are that per color.

\subsection{Freeze-Out}
For $T_{\rm{RH}} > T_{\rm{dec}}$, the relic abundances of $e$ and $u$ are set by freeze-out. 

\toclessnonum\subsection{$b'$ freeze-out}
During the freeze-out of $b$, the decay of $b$ is negligible and we solve the following equation,
\begin{align}
\dot{n}_b + 3Hn_b = -\vev{\sigma_b v_{\rm rel}} (n_b^2 - n_{b,eq}),
\end{align}
$\vev{\sigma_b v}$ is the thermal average of the annihilation cross section times the relative velocity of $b\bar{b}$.
We include the Sommerfeld effect~\cite{Sommerfeld},
\begin{align}
\sigma_q v_{\rm rel} = \frac{2\pi \alpha_{3q,UV}^2}{27 m_q^2 } f (\frac{2\pi c_1 \alpha_{3q,IR}}{v_{\rm rel}}) + \frac{(5 + 6N_{<q} )\pi \alpha_{3q,UV}^2}{27 m_q^2 } f (\frac{2\pi c_8 \alpha_{3q,IR}}{v_{\rm rel}}),\nonumber \\
f(x) = \frac{x}{e^x -1},~c_1 = - \frac{4}{3},~c_8 = \frac{1}{6}, \nonumber \\
\alpha_{3q,UV} \equiv \alpha_3(m_q),~ \alpha_{3q,IR} = \alpha_3(m_q \alpha_3(m_q)),
\label{eq:qannihilate}
\end{align}
where $N_{<q}$ is the total number of quarks and mirror quarks lighter than the mirror quark $q$ (e.g.~$N_{<b}=4$). Here $\alpha_{3q,UV}$ is used for the process with a momentum exchange around the mass of $q$, namely the annihilation, while $\alpha_{3q,IR}$ is used for the process with a momentum exchange around the inverse of the Bohr radius of the $q\bar{q}$ bound state, namely the soft gluon exchange to attract $q\bar{q}$.

\toclessnonum\subsection{$c'$, $\mu'$ and $s'$ freeze-out}
During the freeze-out of $c$, $\mu$ and $s$, the decays of $\mu$ and $s$ are negligible. We solve the following equations,
\begin{align}
\dot{n}_b + 3 H n_b = & - 8 |V_{cb}|^2 \Gamma_b n_b, \\
\dot{n}_c + 3H n_c = &   -\vev{\sigma_c v} (n_c^2 - n_{c,eq}) - 5 \Gamma_c n_c  + 11 |V_{cb}|^2 \Gamma_b n_b, \\
\dot{n}_\mu + 3H n_\mu = &   -\vev{\sigma_\mu v} (n_\mu^2 - n_{\mu,eq}) + 3 |V_{cb}|^2 \Gamma_b n_b + 3 \Gamma_c n_c, \\
\dot{n}_s + 3H n_s = &   -\vev{\sigma_s v} (n_s^2 - n_{s,eq}) + 3 |V_{cb}|^2 \Gamma_b n_b + 5 \Gamma_c n_c,
\end{align}
Here $\Gamma_f$ is defined by
\begin{align}
\Gamma_f = \frac{m_f^5}{1536 \pi^3 v^4}.
\end{align}
The annihilation cross section of a mirror lepton $\ell$ are
\begin{align}
\sigma_\ell v_{\rm rel} = (1 + \sum_{f<\ell} q_f^2 )\frac{\pi \alpha^2}{m_\ell^2} f(-\frac{2\pi \alpha}{v_{\rm rel}}),
\label{eq:lannihilate}
\end{align}
where the summation is taken for mirror fermions lighter than $\ell$ with a charge $q_f$.

\toclessnonum\subsection{$d'$, $u'$ and $e'$ freeze-out}
During the freeze-out of $d$, $u$ and $e$, the decay of $d$ is negligible. The Boltzmann equation is given by
\begin{align}
\dot{n}_\mu + 3H n_\mu = & - 4 \Gamma_\mu n_\mu,\\
\dot{n}_s + 3H n_s = & - 4 |V_{us}|^2 \Gamma_s n_s,\\
\dot{n}_d + 3H n_d = & -\vev{\sigma_d v_{\rm rel}} (n_d^2 - n_{d,eq}) + \Gamma_\mu n_\mu + 3  |V_{us}|^2 \Gamma_s n_s,\\
\dot{n}_u + 3H n_u = & -\vev{\sigma_u v_{\rm rel}} (n_u^2 - n_{u,eq}) + \Gamma_\mu n_\mu + 7  |V_{us}|^2 \Gamma_s n_s,\\
\dot{n}_e + 3H n_e = & -\vev{\sigma_e v_{\rm rel}} (n_e^2 - n_{e,eq}) + \Gamma_\mu n_\mu + 3  |V_{us}|^2 \Gamma_s n_s.
\end{align}
The freeze-out abundance of $d$ is transferred into the abundance of $u$ and $e$ by the mirror beta decay.

\subsection{Freeze-In}

For $T_{\rm{RH}} < T_{\rm{dec}}$, the relic abundances of $e$ and $u$ are set by freeze-in. During the reheating era, the Boltzmann equations are given by
\begin{align}
	\dot{n}_f + 3H n_f =& \vev{\sigma_{HH^\dagger \rightarrow f \bar{f}} \, v_{\rm rel}}(n_H^2 - n_f ^2) + \vev{\sigma_{\rm{therm}}\, v_{\rm rel}}(n_{g}^2 - n_f^2)\Theta(T' - m_f) + ~ \\ &\vev{\sigma_{\rm{therm}}\, v_{\rm rel}}(n_{\gamma}^2 - n_f^2)\Theta(T' - m_f) +  \vev{\sigma_{\rm{f}}\, v_{\rm rel}}(n_{\gamma,eq}^2(m_{f}/T',\mu_\gamma) - n_f^2)\Theta(m_f-T') \nonumber, \\
	\dot{n}_e + 3H n_e =& \vev{\sigma_{HH^\dagger \rightarrow e \bar{e}} \, v_{\rm rel}}(n_H^2 - n_e ^2) + ~ \\ &\vev{\sigma_{\rm{therm}}\, v_{\rm rel}}(n_\gamma^2 - n_e^2)\Theta(T' - m_e) +  \vev{\sigma_{\rm{e}}\, v_{\rm rel}}(n_{\gamma,eq}^2(m_{e}/T',\mu_\gamma) - n_e^2)\Theta(m_e-T')\nonumber, \\
	\dot{n}_{\gamma} + 3H n_{\gamma} =& \vev{\sigma_{HH^\dagger \rightarrow 2{\gamma}} \, v_{\rm rel}}(n_H^2 - n_{\gamma} ^2) + \vev{\sigma_{\rm{2 \rightarrow 3}}\, v_{\rm rel}}(n_f^2 - n_f^2\frac{n_{\gamma}}{n_{\gamma,eq}(T',\mu = 0)}) + ~ 
	\\  &\vev{\sigma_{\rm{therm}}\, v_{\rm rel}}(n_f^2 - n_{\gamma}^2)\Theta(T' - m_f) + \vev{\sigma_{\rm{therm}}\, v_{\rm rel}}(n_e^2 - n_\gamma^2)\Theta(T' - m_e) \nonumber, \\
	\dot{n}_{g} + 3H n_{g} =& \vev{\sigma_{HH^\dagger \rightarrow 2{g}} \, v_{\rm rel}}(n_H^2 - n_{g} ^2) +  \vev{\sigma_{\rm{therm}}\, v_{\rm rel}}(n_f^2 - n_{g}^2)\Theta(T' - m_f)  + ~\\ &\vev{\sigma_{\rm{2 \rightarrow 3}}\, v_{\rm rel}}(n_f^2 - n_f^2\frac{n_{g}}{n_{g,eq}(T',\mu = 0)} + n_g^2 - n_g^2\frac{n_{g}}{n_{g,eq}(T',\mu = 0)}). \nonumber
\end{align}
$f$ is the mirror fermion with the largest mass below $T_{\rm{RH}}$ and subscript $H$ is the SM Higgs.
The production cross sections from the SM Higgs are~\cite{ELLIS1976292,PhysRevD.8.172}
\begin{align}
 \vev{\sigma_{HH^\dagger \rightarrow f \bar{f}} \, v_{\rm rel}} \simeq & \frac{1}{8 \pi} \frac{y_f^2}{v'^2} \\
 \vev{\sigma_{HH^\dagger \rightarrow 2{\gamma}} \, v_{\rm rel}} \simeq &  \frac{1}{16\pi}\left(\frac{\alpha}{4 \pi}\right)^2 \frac{T^2}{v'^4} \left(\sum_f \frac{Q_f^2}{3} \right)^2 \\
 \vev{\sigma_{HH^\dagger \rightarrow 2{g}} \, v_{\rm rel}} \simeq & \frac{1}{2\pi}\left(\frac{\alpha_S}{4 \pi}\right)^2 \frac{T^2}{v'^4}  \left(\sum_q \frac{1}{6} \right)^2,
\end{align}
where the summation on $f$ and $q$ is taken for mirror fermions and quarks with masses greater than $T$.
Initially possessing a typical energy $\sim T$, the thermalization cross-section among mirror charged fermions is given by
\begin{align}
	\vev{\sigma_{\rm{therm}}\, v_{\rm rel}} \approx \frac{4 \pi \alpha_i^2}{{T'}^2}.
	\label{eq:2to2}
\end{align}
while the soft, number-changing $(f \bar{f} \rightarrow f \bar{f} \gamma, \, f \bar{f} \rightarrow f \bar{f}g, \, gg\rightarrow ggg)$ bremsstrahlung cross-sections are given by
\begin{align}
	\vev{\sigma_{\rm{2 \rightarrow 3}}\, v_{\rm rel}} \approx \frac{\alpha_i^3}{2}\left(\frac{\alpha_i n_i}{T'}\right)^{-1} \ln \left(\frac{T'^3}{\alpha_i n_i}\right),
	\label{eq:2to3}
\end{align}
and
\begin{align}
	H = \frac{5}{18} \left(\frac{\pi^2}{10} g_*\right)^{1/2} \frac{T^4}{T_{\rm{RH}}^2 M_{\rm{Pl}}}
\end{align}
is the Hubble scale during the reheating matter-dominated era. Here, $\alpha_i$ equals $\alpha_{\rm EM}$ or $\alpha_S(T')$ and $n_i$ equals $n_e$ or $n_f$ depending on whether the exchange involves mirror photons or gluons.

Soft-scattering keeps the mirror bath in kinetic equilibrium (but not necessarily chemical equilibrium), establishing an effective temperature 
\begin{align}
	T' = \frac{1}{3}\frac{\rho'_{\rm{tot}}(T)}{n'_{\rm{tot}}(T)}
\end{align}
where $\rho'_{\rm{tot}}(T)$ is the total energy density of the mirror sector frozen in via the Higgs portal when the universe is at a temperature $T$, and $n'_{\rm{tot}}$ is the total number density of the mirror sector determined from the Boltzmann equations. For mirror photons, $\gamma$, and gluons, $g$, the equilbrium number densities are
\begin{align}
	n_{eq}\left(\frac{m}{T'}, \,\mu \right) &= g \left(\frac{mT'}{2\pi}\right)^{3/2}\exp \left(-\frac{m}{T'} + \frac{\mu}{T'}\right) = \sqrt{\frac{\pi}{8}}\left(\frac{m}{T'}\right)^{3/2} \exp \left(-\frac{m}{T'}\right) n \\
	n_{eq}(T',\mu = 0) &= \frac{2 g}{\pi^2}T'^3.
\end{align}
For low $v'$ and high $T_{\rm{RH}}$, thermalization of $e$ and $\gamma$ via $2 \to 3$ \eqref{eq:2to3} and $2 \to 2$ \eqref{eq:2to2} processes are effective, thereby increasing $n'_{\rm{tot}}$ and decreasing $T'$. This thermalization acts to cool the mirror bath so that mirror particles freeze-out instantly with an annihilation cross-section $\vev{\sigma_{\rm{f}}\, v_{\rm rel}}$ given by \eqref{eq:qannihilate} if a quark, and \eqref{eq:lannihilate} if a lepton. Nevertheless, these frozen-out particles are then continually replenished by fresh particles from the Higgs portal. Since freeze-in production is maximized at $T_{\rm{RH}}$ and any pre-thermalized contribution is typically small, the most important contributions to the present-day abundance of $e'$ occurs at and below $T_{\rm{RH}}$, discussed below \eqref{eq:tpTRH}-\eqref{eq:boltzmannTRHf}.

For $T < T_{\rm{RH}}$, the universe is radiation dominated. The mirror bath remains in kinetic equilibrium (not necessarily chemical equilibrium), establishing an effective temperature 
\begin{align}
	T' = \frac{1}{3}\frac{\rho'_{\rm{tot}}}{n'_{\rm{tot}}} \simeq \frac{1}{3}\frac{\rho'_{\rm{tot}}(T_{\rm{RH}})}{n_f + n_e + n_\gamma + n_g} \left(\frac{T}{T_{\rm{RH}}}\right)^{4}.
	\label{eq:tpTRH}
\end{align}
The Boltzmann equations for $m_{e} < T' < T_{\rm{RH}}$ determine the evolution of $n_f, n_e, n_g$, and $n_\gamma$, and are given by
\begin{align}
	\dot{n}_f + 3H n_f =& \vev{\sigma_{HH^\dagger \rightarrow f \bar{f}} \, v_{\rm rel}}(n_{H,eq}^2(m_{f}/T) - n_f ^2) + ~ \\ 
	&\vev{\sigma_f \, v_{\rm rel}}(n_{\gamma,eq}^2(m_{f}/T', \,\mu_\gamma) - n_f^2) + \vev{\sigma_f \, v_{\rm rel}}(n_{g,eq}^2(m_{f}/T', \,\mu_g) - n_f^2), \\ 
	\dot{n}_e + 3H n_e =& \vev{\sigma_{HH^\dagger \rightarrow e \bar{e}} \, v_{\rm rel}}(n_H^2 - n_e ^2) +  \vev{\sigma_{\rm{therm}}\, v_{\rm rel}}(n_\gamma^2 - n_e^2), \\
	\dot{n}_\gamma + 3H n_\gamma =& \vev{\sigma_{HH^\dagger \rightarrow 2\gamma} \, v_{\rm rel}}(n_H^2 - n_\gamma ^2) +  \vev{\sigma_f \, v_{\rm rel}}(n_f^2 - n_{\gamma,eq}^2(m_{f}/T', \,\mu_\gamma)) +~ \\ &\vev{\sigma_{\rm{2 \rightarrow 3}}\, v_{\rm rel}}(n_f^2 - n_f^2\frac{n_\gamma}{n_{\gamma,eq}(T',\mu = 0)}) + \vev{\sigma_{\rm{therm}}\, v_{\rm rel}}(n_e^2 - n_\gamma^2)\Theta(T' - m_e). \nonumber \\
	\dot{n}_g + 3H n_g =& \vev{\sigma_{HH^\dagger \rightarrow 2g} \, v_{\rm rel}}(n_H^2 - n_g ^2) +  \vev{\sigma_f \, v_{\rm rel}}(n_f^2 - n_{g,eq}^2(m_{f}/T', \,\mu_g)) +~ \\ &\vev{\sigma_{\rm{2 \rightarrow 3}}\, v_{\rm rel}}(n_f^2 - n_f^2\frac{n_g}{n_{g,eq}(T',\mu = 0)}+ n_g^2 - n_g^2\frac{n_{g}}{n_{g,eq}(T',\mu = 0)}))\nonumber
\end{align}
Last, $e'$ freezes-out when $T'$ drops below  its mass. The Boltzmann equation for $T' < m_{e}$ is 
\begin{align}
	\dot{n}_e + 3H n_e =& \vev{\sigma_{e}\, v_{\rm rel}}(n_{\gamma,eq}^2(m_{e}/T', \,\mu_\gamma) - n_e ^2) 
	\label{eq:boltzmannTRHf}
\end{align}

\section{Energy densities of the mirror QCD bath}
\label{sec:rhos}

In this appendix we estimate the energy density of the mirror QCD bath. We derive the energy density at the phase transition, which is used to estimate the magnitude of gravitational waves, and the energy density of the mirror glueballs after the transition, which is used to estimate the dark radiation abundance. We assume entropy conservation around the mirror QCD phase transition. Entropy production via super-cooling will result in enhancement of the signals.

The SM and mirror sectors decouple from each other at the temperature shown in Eq.~(\ref{eq:Tdec}).
Around this temperature, $e'$, $\mu'$, $u'$, $d'$, $s'$, $g'$, and $\gamma'$ are in the thermal bath; the effective number of degrees of freedom of the mirror sector is $g_{\rm dec}' \simeq 60$. After decoupling, the entropies of the two sectors are separately conserved. Around the mirror QCD phase transition, the mirror gluon bath is nearly pressureless.  Parametrizing the energy density of the mirror gluon bath by $\rho_{g'} = b \, {T'}^4$, the ratio of the temperatures of the two sectors is
\begin{align}
\frac{T_{\rm SM}}{T_{g'}} = 0.3 \left( \frac{g_{\rm dec}}{g_c} \frac{60}{g_{\rm dec}'} \frac{b}{0.5}  \right)^{1/3},
\end{align}
where $g_c$ is the effective number of degrees of freedom of the SM bath at the mirror QCD phase transition. The ratio of the energy densities is
\begin{align}
\frac{\rho_{\rm SM}}{ \rho_{g'}} = 0.5   \left( \frac{106.75}{g_c} \frac{b}{0.5}  \right)^{1/3}  \left( \frac{g_{\rm dec}}{106.75} \; \frac{60}{g_{\rm dec}'} \right)^{4/3}.
\end{align}

For $T' \lesssim 0.7 T'_c$, the energy and the entropy density of the mirror QCD bath is well-approximated by that of the ideal gas of the lightest mirror glueballs with a mass $m_{S'}\simeq 5.3 T_c'$~\cite{Borsanyi:2012ve}.
Entropy conservation within this decoupled mirror bath implies its entropy density scales as $\propto a^{-3}$. $3 \to 2$ annihilations keep warm the mirror glueballs so that
their temperature falls approximately as $\propto \ln a$ and energy density as $\propto a^{-3} \,(\ln{a})^{-1}$ until they decouple or decay~\cite{Carlson:1992fn,Hochberg:2014dra,Forestell:2016qhc}. Here, $a$ is the scale factor of the universe. The $3 \to 2$ cross-section is given by~\cite{Forestell:2016qhc}
\begin{align}
	\langle\sigma_{3 \to 2}v^2 \rangle \simeq \frac{B}{(4 \pi)^3}\left(\frac{4 \pi}{3}\right)^6\frac{1}{m_{S'}^5},
\end{align}
where $B$ is an $\mathcal{O}(1)$ number whose value weakly affects $a_f$. We take $B = 1$. 

As discussed in Sec.~\ref{sec:BBNandDR}, the non-trivial dynamics around the mirror QCD phase transition are encoded in the modification factor $A$, the ratio of the actual mirror glueball energy density to that derived by a non-interacting ideal gas approximation and the glueball number conservation,
\begin{align}
	A = \frac{4 T_f'}{3 T_c'} = \frac{4}{3} \frac{2 m_{S'}}{T_c'} W\left(\frac{2}{(2 \pi)^3} \left(\frac{45}{32 \pi^2}\right)^2 \left( \frac{m_{S'}}{T_c'} \, \frac{a_f}{a_c}\right)^6\right)^{-1} \appropto \left( \ln\frac{a_f}{a_c}\right)^{-1}.
	\label{eq:gBallFudge}
\end{align}
Here, $W(x)$ is the product-log function, which is a solution of $W e^W = x$. $a_c$ is the would-be scale factor at $T'= T_c'$ if the mirror gluons remain an ideal gas until the phase transition, and $a_f$ is the scale factor of the universe when the $3 \to 2$ reactions among mirror glueballs freeze-out, or the mirror glueballs decay. For $v' > 10^9$ GeV, $a_f$ is determined by the former and otherwise by the latter.

\begin{figure}[tb]
\centering
\includegraphics[width=0.7\textwidth]{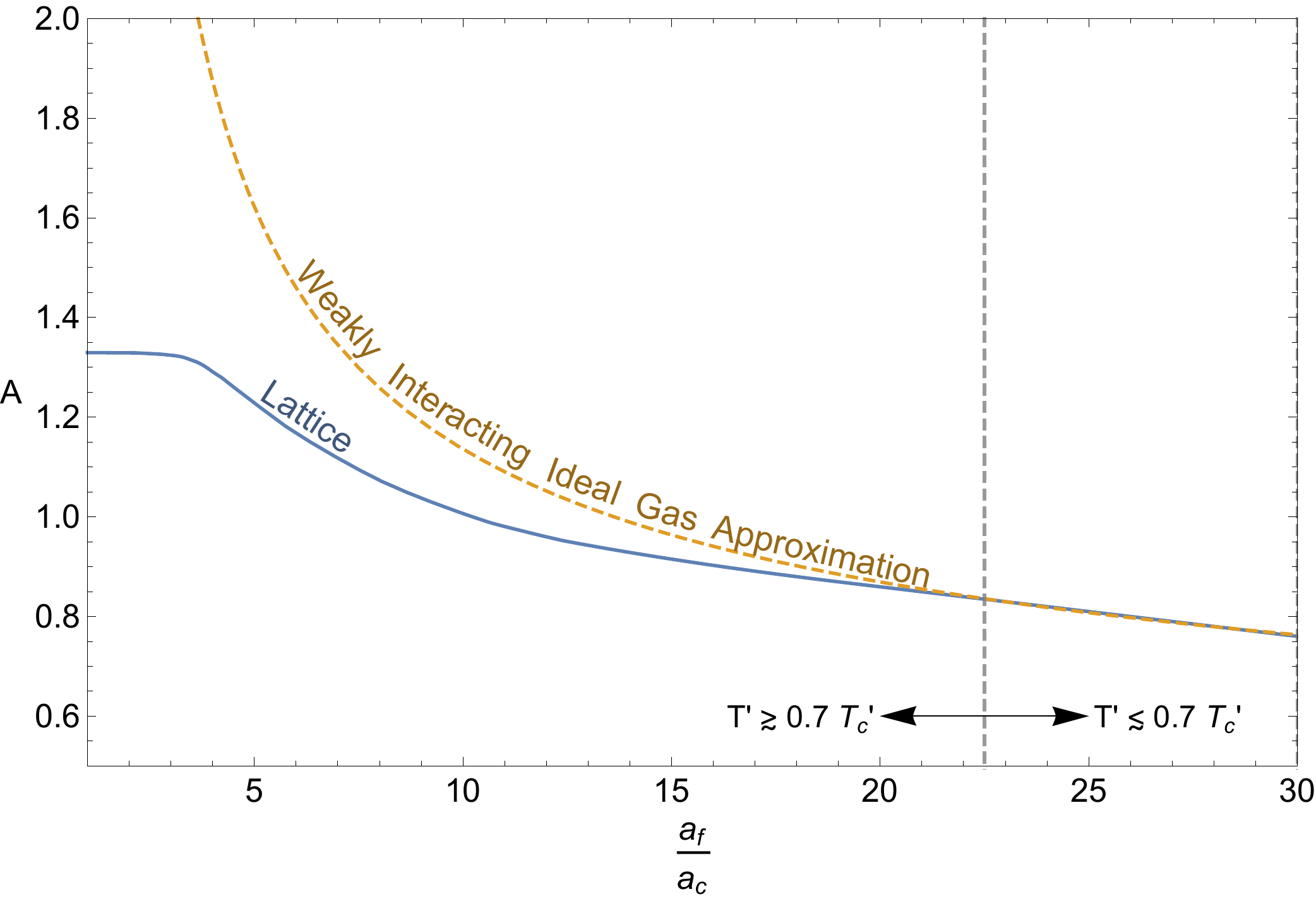} %
\caption{\small The QCD$'$ modification factor $A$ as a function of $a_f/a_c$. $A$ is defined as the energy density ratio of the actual glueball gas to that derived by a non-interacting ideal gas approximation and glueball number conservation.}
\label{fig:aFactor}
\end{figure}

For $0.7 T'_c \lesssim T' \lesssim T'_c$, the energy density of the mirror glueball bath deviates from that of a weakly-interacting ideal gas composed of the lightest mirror glueballs, and hence the second equality of $\eqref{eq:gBallFudge}$ is invalid. In this strongly interacting regime, $A$ is determined by taking the lattice result for $\rho_{g'}(T'_f/T'_c)$ from~\cite{Borsanyi:2012ve} and equating it with $s_{g'} T'_f$ - an excellent approximation since the glueball gas is nearly pressureless. Here, $s_{g'} = 32 \pi^2/45 {T'_c}^3 (a_c/a_f)^3$ is the entropy density of the mirror glueball bath. $T'_f/T'_c$ is then numerically solved for as a function of $a_f/a_c$ and inserted into $\eqref{eq:gBallFudge}$ to determine $A$ as function of $a_f/a_c$ as shown for both regimes in Fig.~\ref{fig:aFactor}.

\bibliography{mirrorDM} 

\let\addcontentsline\oldaddcontentsline
  
\end{document}